\input harvmac
\input epsf.tex
\overfullrule=0mm
\newcount\figno
\figno=0
\def\fig#1#2#3{
\par\begingroup\parindent=0pt\leftskip=1cm\rightskip=1cm\parindent=0pt
\baselineskip=11pt
\global\advance\figno by 1
\midinsert
\epsfxsize=#3
\centerline{\epsfbox{#2}}
\vskip 12pt
{\bf Fig. \the\figno:} #1\par
\endinsert\endgroup\par
}
\def\figlabel#1{\xdef#1{\the\figno}}
\def\encadremath#1{\vbox{\hrule\hbox{\vrule\kern8pt\vbox{\kern8pt
\hbox{$\displaystyle #1$}\kern8pt}
\kern8pt\vrule}\hrule}}
\def\appendix#1#2{\global\meqno=1\global\subsecno=0\xdef\secsym{\hbox{#1.}}
\bigbreak\bigskip\noindent{\bf Appendix. #2}\message{(#1. #2)}
\writetoca{Appendix {#2}}\par\nobreak\medskip\nobreak}

\def\tvi{\vrule height 12pt depth 6pt width 0pt}
\def\tv{\tvi\vrule}

\Title{T95/019, SU-4240-602, cond-mat/9502063}
{{\vbox {
\bigskip
\centerline{{\bf Three--dimensional Folding}}
\centerline{{\bf of the Triangular Lattice}}
}}}
\bigskip
\centerline{M. Bowick,}
\medskip
\centerline{\it Physics Department, Syracuse University,}
\centerline{\it Syracuse NY 13244-1130, USA}
\medskip
\centerline{P. Di Francesco, O. Golinelli and E. Guitter}
\medskip
\centerline{ \it Service de Physique Th\'eorique de Saclay,}
\centerline{ \it F-91191 Gif sur Yvette Cedex, France}
\baselineskip=24pt
\vskip .5in
 
We study the folding of the regular triangular lattice in three dimensional embedding space,
a model for the crumpling of polymerised membranes. 
We consider a discrete model, where folds are either planar or form the angles
of a regular octahedron. These ``octahedral" folding rules correspond simply to a discretisation
of the $3d$ embedding space as a Face Centred Cubic lattice. The model is shown to be equivalent
to a 96--vertex model on the triangular lattice. The folding entropy per triangle
${\rm ln}\ q_{3d}$ is evaluated numerically to be $q_{3d}=1.43(1)$. Various exact bounds
on $q_{3d}$ are derived.

\noindent
\Date{02/95}
 

\nref\KN{Y. Kantor and D.R. Nelson, Phys. Rev. Lett. {\bf 58} (1987) 2774
and Phys. Rev.  {\bf A 36} (1987) 4020.}
\nref\NP{D.R. Nelson and L. Peliti, J. Phys. France {\bf 48} (1987) 1085.}
\nref\PKN{M. Paczuski, M. Kardar and D.R. Nelson, Phys. Rev. Lett. {\bf 60} (1988) 2638.}
\nref\DG{F. David and E. Guitter, Europhys. Lett. {\bf 5} (1988) 709.}
\nref\TAS{M. Baig, D. Espriu and J. Wheater, Nucl. Phys. {\bf B314} (1989) 587;
R. Renken and J. Kogut, Nucl. Phys. {\bf B342} (1990) 753; R. Harnish and
J. Wheater, Nucl. Phys. {\bf B350} (1991) 861; J. Wheater and P. Stephenson, Phys. Lett. 
{\bf B302} (1993) 447.}
\nref\KJ{Y. Kantor and M.V. Jari\'c, Europhys. Lett. {\bf 11} (1990) 157.}
\nref\DIG{P. Di Francesco and E. Guitter, Europhys. Lett. {\bf 26} (1994)
455.}
\nref\BAX{R.J. Baxter, J. Math. Phys. {\bf 11} (1970) 784 and
J. Phys. A: Math. Gen. {\bf 19} (1986) 2821.}
\nref\DIGG{P. Di Francesco and E. Guitter, Phys. Rev. E {\bf 50} (1994) 4418.}
\nref\SHAN{D. Shanks, J. Math. Phys. {\bf 34} (1955) 1.}
\nref\BREZ{C. Brezinski and M. Redivo Zaglia, {\it Extrapolation methods}
(Elsevier North-Holland, 1991).}
\nref\NIEN{E. Domany, D. Mukamel, B. Nienhuis and A. Schwimmer, Nucl. Phys.
{\bf B190} (1981) 279; B. Nienhuis, Phys. Rev. Lett. {\bf 49} (1982) 1062
and J. Stat. Phys. {\bf 34} (1984) 731.}


Polymerised membranes have recently been the object of intense investigation,  both
as natural $2d$ generalisations of polymers and as models for biological systems. 
{}From the theoretical point of view, polymerised membrane models are archetypical examples
of the interplay between geometry and statistical mechanics.
In this paper, we consider only phantom membranes, where steric constraints
due to self--avoidance are not taken into account.
As in other domains of random surface physics, the recourse to discrete formulations provides
a powerful approach, with the advantage of permitting direct numerical simulations, as well as
leading to exact solutions for particular models. 
In the simplest discretised version, the polymerised membrane is modelled by a regular
triangular lattice with fixed connectivity, embedded in a $d$--dimensional space. 
In one class of models, the lengths of the links of the lattice are allowed to
have small variations. These {\it tethered} membranes were first studied in \KN, 
where a geometrical $3d$ continuous crumpling transition was predicted. This prediction
was corroborated by various analytic results on continuous models [\xref\NP -\xref\DG]
and other numerical simulations \TAS.
In a second class of models, the lengths of the links of the lattice are fixed, 
say to unity. For such membranes, the only remaining degree of freedom is  that
of {\it folding} of the lattice, whose links serve as hinges between neighbouring triangles.
Membrane folding was first studied in $2d$ embedding space, where it can be
formulated as an 11--vertex model on the triangular lattice. 
The entropy of folding, which counts the number of distinct folded states of the lattice
in the thermodynamic limit, was first estimated numerically in \KJ. The $2d$ folding
problem was then solved exactly in \DIG, through the equivalence with 
the 3--colouring problem of the links of the triangular lattice \BAX.  
A further study, including a bending rigidity, performed in \DIGG, led to numerical
evidence of a first order folding transition of the membrane.

In this paper, we address the $3d$ folding problem, where the embedding space now
has 3 dimensions, in an attempt to recover results for tethered membranes in the
folding language.  In particular the $2d$ folding transition could be smoothed and
lead to a continuous $3d$ folding transition, analogous to the crumpling transition.
The present work is devoted to a suitable definition of the $3d$ folding problem,
amenable to numerical simulations, and to various reformulations as 
a vertex model, constrained spin system and colouring problem.  
We use these formulations to estimate the entropy of $3d$ folding both numerically and 
analytically.

The paper is organised as follows. In the first section, we define the general problem of folding
of the triangular lattice in embedding space with arbitrary dimension and recall a few
results for the case of dimension 2.
In section 2, we define the discrete problem of $3d$ ``octahedral" folding, 
which corresponds to a discretisation of the $3d$ embedding space as  
a Face Centred Cubic lattice. In section 3, we show that the $3d$ octahedral folding is
equivalent to a 96--vertex model on the triangular lattice. This is done by use of face spin variables
defined in section {\it (3.1)}, which in turn parametrise some generators of the 
tetrahedral group studied in section {\it (3.2)}. 
{}From the group formulation we derive in section {\it (3.3)} the local folding constraints on these
face variables, leading to a complete determination of the 96 folding vertices.
Section 4 is devoted to the numerical evaluation of the entropy of $3d$ folding.
Various exact bounds on this entropy are derived in section 5, by reexpressing the $3d$ octahedral
folding problem as a dressed 3--colouring problem (section {\it (5.1)}), and performing some
rough bounding of the partition function (section {\it (5.2)}). Improved bounds
are obtained in section {\it (5.3)} by comparison with the $2d$ folding problem in a field.
We discuss in section 6 our results for the entropy (section {\it (6.1)}), as
well as a natural generalisation of the octahedral folding problem in dimension $d>3$ (section {\it (6.2)}).
Concluding remarks are gathered in section 7, and some extensions of section {\it (5.2)} are 
given in the Appendix.

\newsec{The folding rule}
 
A folding in ${\bf R}^d$ of the regular triangular lattice is a mapping  
which assigns to each vertex $v$ of the triangular lattice a position ${\bf X}_v$ in 
$d$-dimensional embedding space ${\bf R}^d$, with the ``metric'' constraint that the
Euclidean distance $|{\bf X}_{v_2}-{\bf X}_{v_1}|$ in ${\bf R}^d$ between nearest neighbours 
$v_1$ and $v_2$ on the lattice is always unity. 
Under such a mapping, each elementary triangle of the lattice is mapped onto 
an equilateral triangle in ${\bf R}^d$. 
In general, two adjacent triangles form some angle in ${\bf R}^d$, i.e.
links serve as hinges between triangles and may be (partially) folded.
\fig{The oriented triangular lattice: triangles pointing up (resp. down) are
oriented counterclockwise (resp. clockwise). The three tangent vectors ${\bf t}_i$,
$i=1,2,3$, have a vanishing sum in the embedding space.}{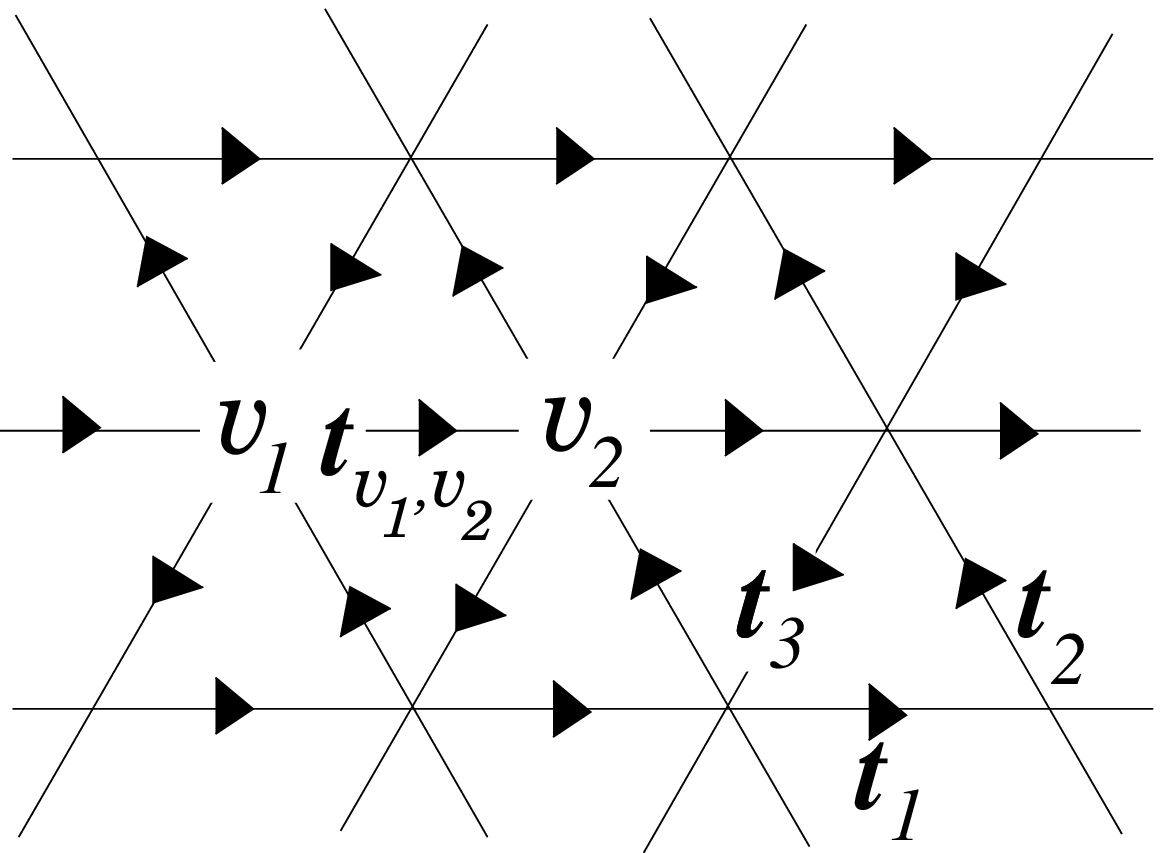}{2.0truein}
\figlabel\lattice
\noindent Folding is best described in terms of tangent vectors, which are link
variables defined as follows:
we first orient the links of the lattice as in Fig.\lattice, with triangles 
pointing up (resp. down) oriented counterclockwise (resp. clockwise), and define 
the tangent vector between two neighbours $v_1$ and $v_2$ as the vector 
\eqn\tangent{{\bf t}_{v_1,v_2}= {\bf X}_{v_2}-{\bf X}_{v_1}}
if the arrow points from $v_1$ to $v_2$. The metric constraint states
that all tangent vectors have {\it unit length}.
Moreover, with our choice of orientation, the three tangent vectors
${\bf t}_i$, $i=1,2,3$, around each face of the lattice must
have {\it vanishing sum}. This is the basic folding rule\foot{On the
dual hexagonal lattice, whose vertices are at the centers of the triangles, 
this basic rule translates into a local conservation law at each vertex,
reminiscent of the so--called ``ice rule" of integrable vertex models.}:
\eqn\foldrule{{\bf t}_1 +{\bf t}_2+ {\bf t}_3 = {\bf 0}.}
Up to a global translation in ${\bf R}^d$, a folding is therefore a
configuration of unit tangent vectors defined on the links of the lattice, 
obeying the folding rule \foldrule\ around each triangle.

The two--dimensional ($d=2$) folding problem of the triangular lattice 
was addressed in \DIG. It is easy to check that, up to a global rotation 
in the embedding plane, all the link variables are forced to take their value 
among a {\it fixed} set of three unit vectors with vanishing sum. This permits 
a reformulation of the $d=2$ folding problem as that of the 3--colouring of 
the links of the triangular lattice: calling the three fixed vectors blue,
white and red, the folding rule translates into the constraint
that the three colours around each triangle have to be distinct. This 
3--colouring problem was solved by Baxter in \BAX\ by use of
Bethe Ansatz techniques. His result for the thermodynamic partition
function measures the number of folding configurations $Z_{2d}\propto q_{2d}^{N_\Delta}$
for a lattice with $N_\Delta$ triangles, 
in the limit of large $N_\Delta$.
This gives the folding entropy per triangle $s_{2d}=\log (q_{2d})$, with  \DIG\
\eqn\exatwod{ q_{2d}~=~ {\sqrt{3} \over 2 \pi} \Gamma(1/3)^{3/2}~=~
1.20872... }
The $2d$ folding problem has also been studied in the presence of bending
rigidity, which associates an energy to each folded link, and a 
magnetic field coupled to the normal vector to the triangles \DIGG.
The system was found to undergo a first order folding transition.
At zero rigidity and zero magnetic field, the lattice is in an entropic folded
phase. At large enough rigidity and/or magnetic field, the lattice becomes
totally unfolded.

In this paper, we address the case of embedding in ${\bf R}^{d=3}$
and, in particular, we are interested in computing the folding entropy $s_{3d}$. 

\newsec{Octahedral Folding}

In the general $3$--dimensional folding problem,
the local folding constraint \foldrule\ imposes only
that the three tangent vectors around each face be in the same plane
and have relative angles of $2 \pi/3$. This, however, does not
impose any constraint on the {\it relative} positions of the two planes
corresponding to two adjacent faces, which may form some arbitrary 
continuous angle. As opposed to the $2d$ case, this then leads 
to a problem with continuous degrees of freedom.

Here we wish instead to define some {\it discrete} model of folding in $3d$, 
in which only a {\it finite} number of relative angles are allowed between 
adjacent faces. More specifically, we shall also impose that the link variables
themselves
take their values among a {\it finite} set of tangent vectors, now in ${\bf R}^3$.
For symmetry reasons, we will take this set of tangent vectors
to be the (oriented) edges of a regular solid of ${\bf R}^3$, made of equilateral 
triangles only.  

\fig{The oriented octahedron: the edges around each face form triplets of 
tangent vectors with vanishing sum. The four normal vectors ${\bf n}_i$, $i=1,2,3,4$,
are represented on the corresponding outward oriented faces.}{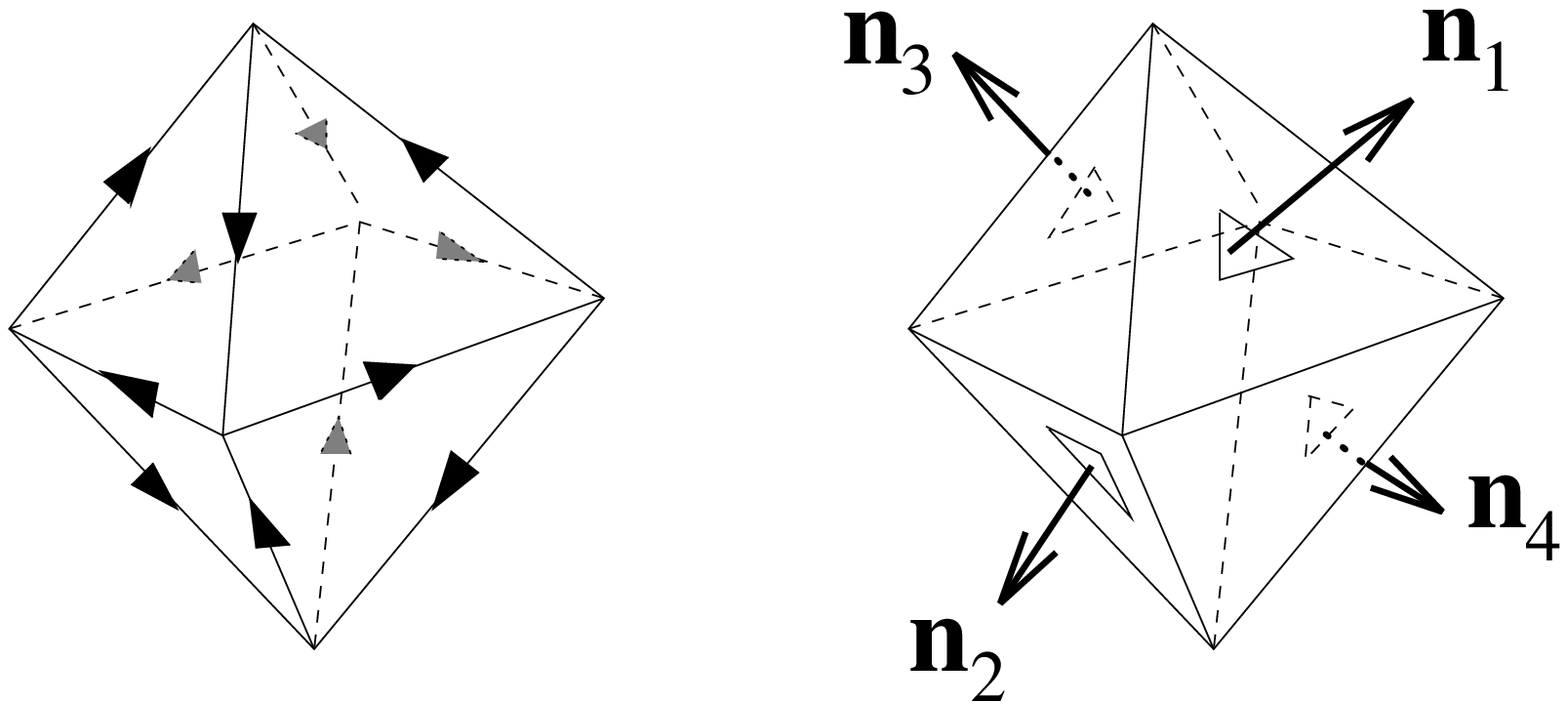}{8.truecm}
\figlabel\octaor

\noindent There are only three regular solids in ${\bf R}^3$ made of equilateral triangles:
the tetrahedron, the octahedron and the icosahedron. 
The edges of the tetrahedron (resp. icosahedron), however, cannot be consistently oriented 
in order for the corresponding tangent vectors to satisfy \foldrule\ around each face.
This is because each vertex is surrounded by an odd number 3 (resp. 5) of triangles.  
There is no such problem for the octahedron, as shown in Fig.\octaor.
The $12$ links of the octahedron are oriented consistently to form $8$
triplets of tangent vectors with vanishing sum corresponding to the $8$ faces
of the octahedron.  
{}From now on, we shall therefore consider the restricted $3d$ ``octahedral folding'' 
problem, where the tangent vectors are chosen from the set of the 12 edge vectors of a
regular oriented octahedron. In the folding process, the folding rule \foldrule\ 
imposes that
the three links of a given face on the original triangular lattice are mapped onto 
one of the 8 triplets of tangent vectors above. For a given triplet, the triangle
can still be in $3!$ states corresponding to the $3!$ permutations of the three edges.  
Each triangle can therefore be in one of $48=8\times 6$ states.

The 8 faces of the octahedron can be labelled as follows:  
we consider for each face its normal vector, pointing outwards or inwards according to the 
orientation of its tangent vectors on the octahedron (see Fig.\octaor). 
There are four outward oriented and four inward oriented faces which alternate 
on the octahedron. The normal vectors to opposite faces are equal. This thus defines 
a set of only four vectors ${\bf n}_1$, ${\bf n}_2$, ${\bf n}_3$ and ${\bf n}_4$,
furthermore satisfying the sum rule 
${\bf n}_1+{\bf n}_2+{\bf n}_3+{\bf n}_4={\bf 0}$.
Each face is labelled by its orientation (outward or inward) and its normal vector 
(1, 2, 3 or 4).
Notice that, by labelling the normal vectors, we can arbitrarily fix one of the 
two possible chiralities of the octahedron in this construction. We choose the one 
corresponding to Fig.\octaor.
\fig{The Face Centred Cubic lattice viewed as a packing of $3d$ space with octahedra and
tetrahedra.}{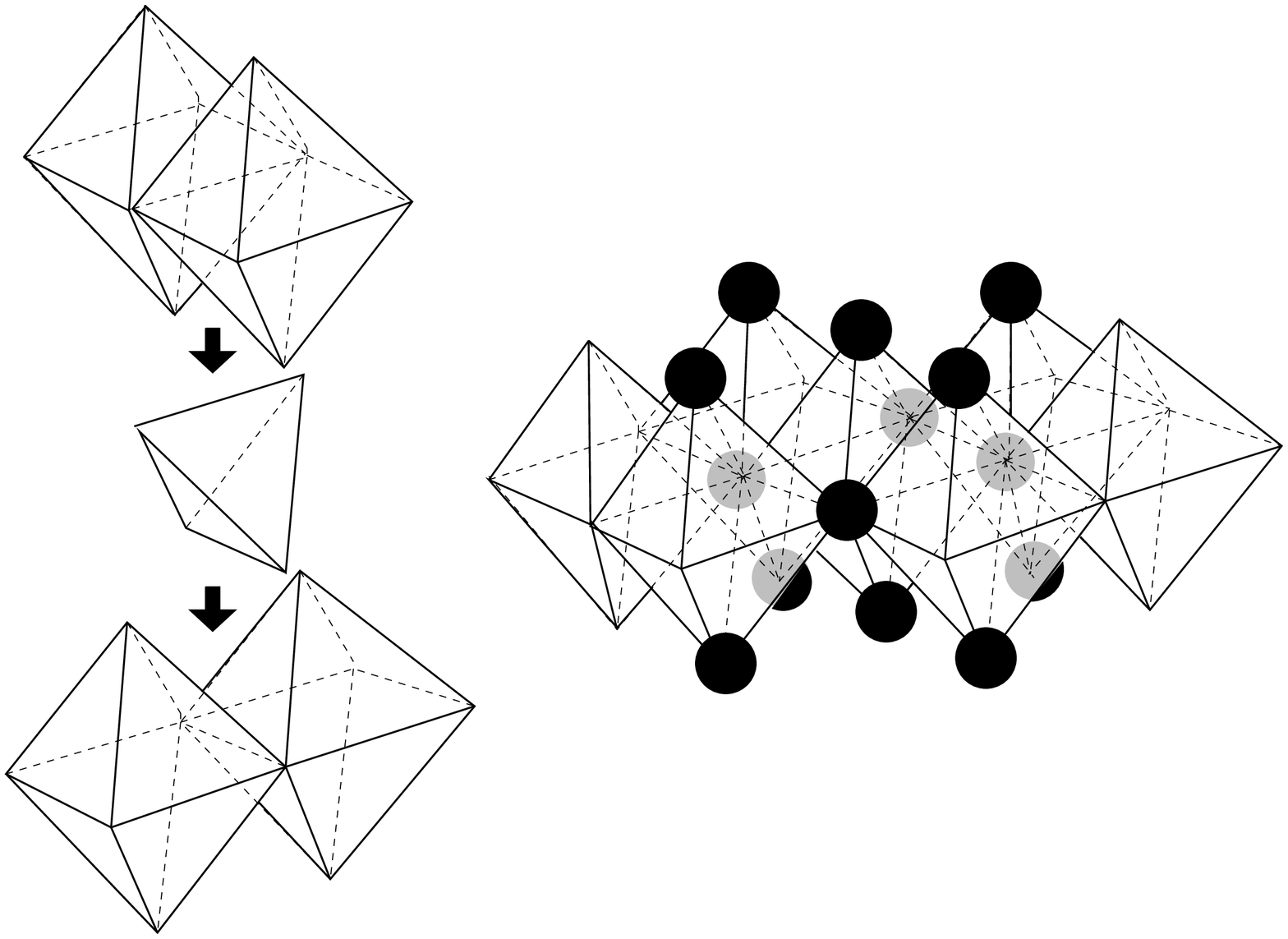}{4.0truein}
\figlabel\fcc

The 12 oriented edge vectors of the octahedron are actually identical to the
6 edge vectors of a tetrahedron, now taken with the two possible orientations.   
The four normal vectors above are also the normals to this tetrahedron.
For each folding map, the image of the folded lattice in ${\bf R}^3$ 
lies therefore on a 3d Face Centred Cubic (FCC) lattice, which consists of a 
filling of space by octahedra complemented by tetrahedra, as shown in Fig.\fcc.  
In this respect, the ``octahedral folding'' problem simply corresponds to
discretising the embedding space as a FCC lattice.

Note finally that nothing prevents the lattice from intersecting itself,
hence our construction describes a phantom membrane. 
The introduction of self--avoidance
would result in much more elaborate non--local constraints, 
far beyond the scope of the present study.

\newsec{96--vertex Model}

When stated in terms of tangent vectors, the $3d$ ``octahedral folding'' problem
involves three types of constraints: face, link and vertex constraints. 
The first constraint, around each {\it face}, imposes 
that the three tangent vectors of a given triangle form one of the $8\times 6$
(ordered) triplets with vanishing sum. The second constraint, on each {\it link}, arises
because two adjacent triangles share a common tangent vector. Given the
state of one triangle, any adjacent triangle has one of its tangent vectors 
already fixed and thus is left with only $4=48/12$ possible states. 
\fig{The four possible folding angles between two adjacent triangles. The neighbour
of the dark triangle may (i) be itself on top of the dark triangle (complete fold), 
(ii) occupy the symmetric position in the same plane (no fold), (iii) lie on the 
same octahedron (i.e. form an obtuse angle) or
(iv) lie on the same tetrahedron (i.e. form an acute angle).}{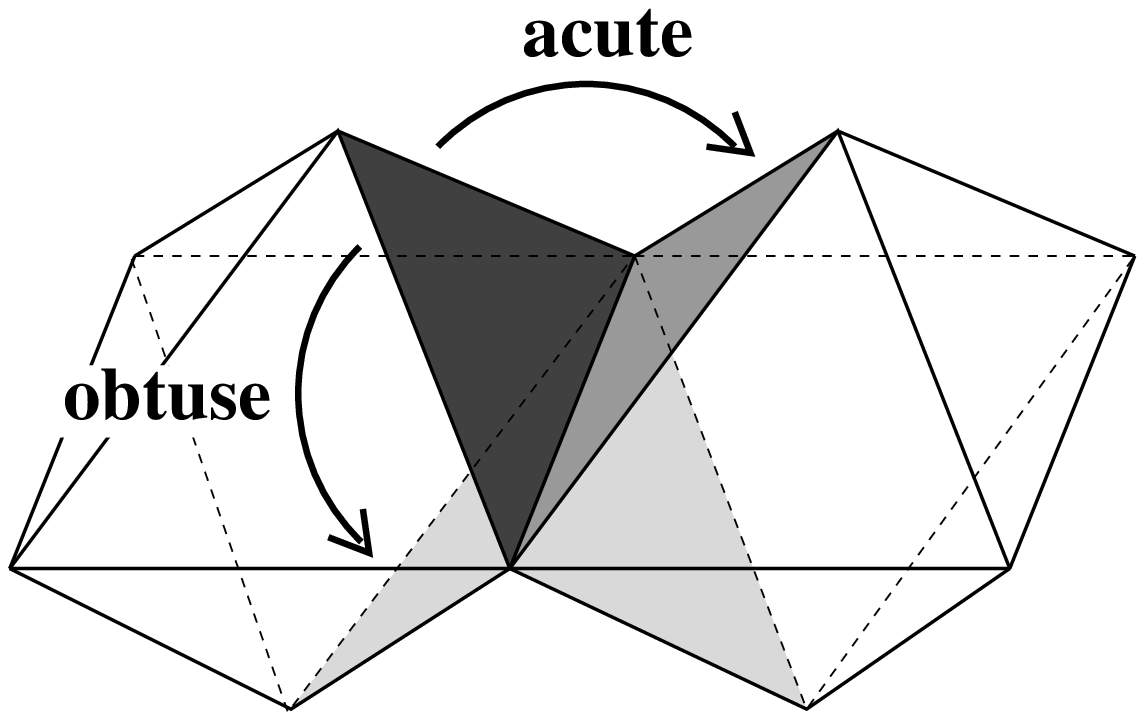}{6.truecm}
\figlabel\acutob
\noindent They
correspond simply to the four values for the relative angle between two
neighbouring triangles, i.e. the angle between the normal vectors, depicted in Fig.\acutob.
These four values are 0 (no fold: the triangles are side by side), 
$180^o$ (complete fold: the triangles are on top of each other), 
$\arccos(-1/3)\sim 109^o28'$ (fold with acute angle: the two triangles lie on the same
tetrahedron) and $\arccos(1/3)\sim 70^o32'$ (fold with obtuse angle: the triangles lie
on the same octahedron). 
Finally, there is a third constraint on the six successive folds 
around each 
{\it vertex} of the lattice:  after making one loop,
the same tangent vector must be recovered. 
Since the ``metric constraint'' is local,
there are actually no constraints other than these three (face, link and vertex)
constraints. 
  
In the study of the $2d$ folding problem, i.e. of the 3--colouring problem,
the face and link constraints are taken into account by going to ${\bf Z}_2$ spin
variables $\sigma_i$ defined on the faces of the lattice. Ordering the colours 
cyclically, the spin is $+1$ (resp.  $-1$) if the colour increases (resp. decreases) 
from one link to the neighbouring one on the triangle, oriented counterclockwise. 
In this language, the actual folds take place exactly on the domain walls of the
spin variable. Instead of having a ${\bf Z}_3$ colour variable per link, one is left with
a ${\bf Z}_2$ spin variable per triangle. The vertex constraint translates into a constraint
on the six spins $\sigma_1,...,\sigma_6$ around each vertex of the lattice, namely
that $\sum_{i=1}^6\sigma_i=0$ mod 3. This leads to 22 possible local spin configurations 
around each vertex, or equivalently, after removing the global ${\bf Z}_2$ degeneracy of
reversal of all spins, to an 11--vertex model on the lattice \DIG.

In this section, we shall proceed in the same way for the $3d$ ``octahedral folding''
and account for the face and link constraints by expressing folded configurations
in terms of two ${\bf Z}_2$ variables on the triangles. 
These variables will indicate the relative states of successive links around the face.
We shall then count the number
of allowed hexagonal configurations around a vertex: we will find that our problem
is now expressible as a 96--vertex model. These vertices and the corresponding rules
on the ${\bf Z}_2$ variables will be identified in the next section.

\subsec{Face variables and counting of the vertices}

\fig{The labelling $(i,j)$ of an edge of the octahedron, $1 \leq i \neq j \leq 4$, 
according to the adjacent normal vectors.}{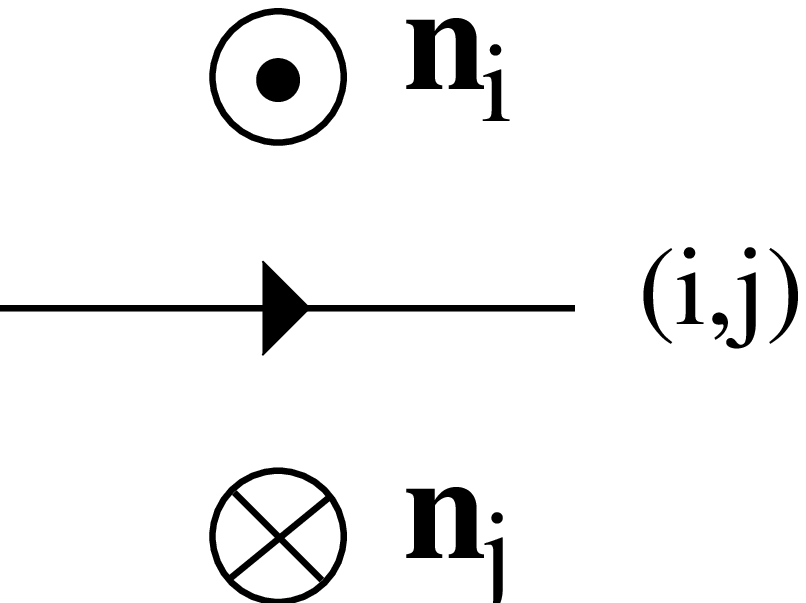}{1.5truein}
\figlabel\tanlab

Let us label the $12$ edges of the octahedron as follows:
each edge is shared by two adjacent faces, one outward and one inward oriented (see Fig.\octaor).
We label the edges by the indices $(i,j)$, $1 \leq i \neq j \leq 4$, 
when the normal vector for the outward face is ${\bf n}_i$ and the one for the inward
face is ${\bf n}_j$. This convention is illustrated in Fig.\tanlab.
There are $12$ such couples $(i,j)$.
The unit tangent vector associated with the link $(i,j)$
is given by
\eqn\tij{{\bf t}_{(i,j)}= {3 \over 2 \sqrt 2} \, {\bf n}_i 
\times (- {\bf n}_j).}

\fig{The transition from a link $(i,j)$ to a subsequent link $(k,l)$ is described by
the two ${\bf Z}_2$ face variables $z$ and $\sigma$.}{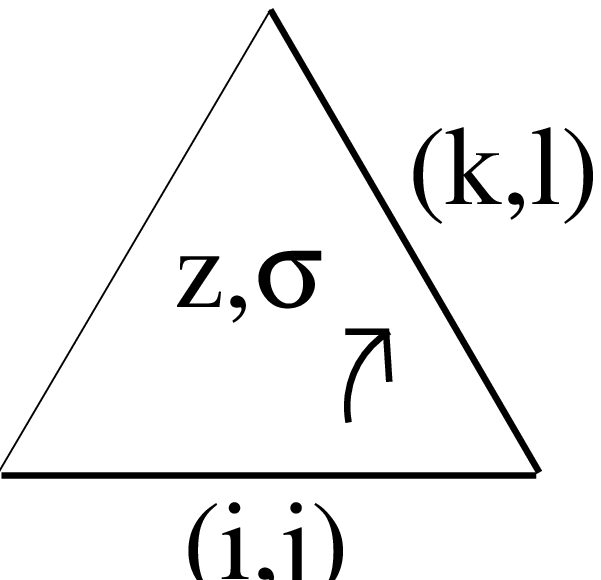}{3.truecm}
\figlabel\onestep

Consider now an elementary triangle of the lattice. 
Starting from one of its links $(i,j)$ the subsequent link $(k,l)$ counterclockwise
must share a face with $(i,j)$ on the octahedron.
This leads to the $4$ following possibilities, labelled
by the ${\bf Z}_2$ face variables $z, \ \sigma \in \{\pm 1\}$:
\eqn\fourchoice{
\eqalign{& z=+1: (i,j) \to (i,l), \ l \neq j, \quad 
\epsilon_{ijl} = - \sigma= \pm 1 \cr
& z=-1: (i,j) \to (k,j), \ k\neq i, 
\quad \epsilon_{ijk} = + \sigma= \pm 1 \cr}}
where $\epsilon_{ijk}=\sum_l \epsilon_{ijkl}$ is defined
in terms of the totally antisymmetric tensor $\epsilon_{ijkl}$, equal to the signature of the
permutation $(ijkl)$ of $(1234)$.  The value
$z=+1$ (resp. $z=-1$) indicates that the two tangent vectors share
an outward oriented (resp. inward oriented) face on the octahedron. 
The spin variable
$\sigma$ takes the value $+1$ (resp. $-1$) if $(k,l)$ follows (resp. precedes) 
$(i,j)$ on their common (oriented) face of the octahedron. 
Using \tij, one can check that the variable $\sigma$ also indicates whether 
the normal vector to the triangle (in the embedding space ${\bf R}^3$) is parallel ($\sigma=+1$)
or antiparallel ($\sigma=-1$) to the corresponding normal vector of the octahedron. 

Considering now two neighbouring triangles, the $4$ possible relative values
$z_2/z_1$ and $\sigma_2/\sigma_1$ indicate which type of fold they form, 
with the correspondence displayed in Table I below.
%
$$\vbox{\offinterlineskip
\halign{\tv\quad # & \quad\tv \quad
# & \quad \tv \quad  # & \quad \tv #\cr
\noalign{\hrule}
\tvi $z_2/z_1$ & $\sigma_2/\sigma_1$ & angle &\cr
\noalign{\hrule}
\tvi  $\phantom{-}1$ &  $\phantom{-}1$  &    no fold  &\cr
\tvi  $\phantom{-}1$ &  $-1$  &    complete fold  &\cr
\tvi  $-1$ &  $\phantom{-}1$  &    acute fold  &\cr
\tvi  $-1$ &  $-1$  &    obtuse fold  &\cr
\noalign{\hrule} }} $$
\par\begingroup\parindent=0pt\leftskip=1cm\rightskip=1cm\parindent=0pt
\baselineskip=11pt
{\bf Table I:} 
The relative folding state of two neighbouring triangles
according to their relative values of $z$ and $\sigma$.
\par
\endgroup\par
\noindent The domain walls for the $z$ variable are the location of the 
folds which are either acute
or obtuse, whereas those for the $\sigma$ variable are the location of the folds which are
either complete or obtuse. 
The superposition of these two types of domain walls fixes the folding state of all the links,
specifying the folding state of
the lattice up to a global orientation.

The use of $z$ and $\sigma$ variables instead of the $12$ $(i,j)$ variables incorporates
the face and link constraints. Like in the $2d$ case, the vertex constraint is more
subtle and will be studied in the next section.
Nevertheless, we can easily count at this stage the number of possible configurations around a vertex 
satisfying this constraint, i.e. the number of possible folded states of an elementary hexagon.
Indeed the mapping \fourchoice\ may be represented by a $12\times 12$ connectivity matrix
$M_{(i,j),(k,l)}$ with $i\ne j$ and $k\ne l$:
\eqn\connectivity{M_{(i,j),(k,l)}= \delta_{ik}+\delta_{jl}-2\, \delta_{ik}\delta_{jl} \ .}
This matrix acts as a transfer matrix between two successive internal links of the hexagon.
The number of configurations of a hexagon is simply given by:
\eqn\vert{\Tr (M^6)~=~4608 \ ,}
where the trace guarantees that the same link variable is recovered after one loop.
These $4608$ configurations count as distinct all the foldings which are
related by a global change of 
orientation of the hexagon in embedding space. 
The order of the resultant degeneracy is 
$48$, corresponding to $12$ choices for the first tangent on the octahedron
times $4$ for the choice of the second from among its 4 neighbours  
(this latter choice corresponds to the $4$ choices
of the $z$ and $\sigma$ variables on the corresponding triangle).
This leaves us with $4608/48 = {\bf 96}$ distinct configurations.

The above computation is also equivalent to counting the number of closed paths 
of length $6$ on the cuboctahedron, i.e. the solid whose vertices sit
at the centers of the edges of the octahedron, and whose connectivity
matrix is given by \connectivity.

\subsec{The tetrahedral group $A_4$}

In order to derive the vertex constraints for the $z$ and $\sigma$ variables around
any vertex of the lattice, we will use a group formulation of \fourchoice.
We note here that the $12$ links of the octahedron may be uniquely labelled
by the even permutations of the set $1,2,3,4$. Indeed the link 
$(i,j)$ is equivalently represented by the even permutation 
$(ijkl)$, with $\epsilon_{ijkl}=1$.
With this labelling the alternating group $A_4$, the group of even
permutations of four elements, also known as the tetrahedral group, 
acts on the space of links. 
In other words, each element of $A_4$ is a one-to-one
mapping of the set of links onto itself. A fixed element of $A_4$ acts
on a given link by permutation of the four link labels. 
The order of $A_4$ is $4!/2=12$. 
These $12$ elements map a given link to exactly the $12$ links of the octahedron.
The group $A_4$ can be generated by two elements $\tau_1$ and $\tau_3$ defined as:
\eqn\perm{\eqalign{\tau_1 ~&=~ (243) \cr
\tau_3 ~&=~ (134) \cr }}
\fig{The action of the generators $\tau_1$ and $\tau_3$ of the tetrahedral group $A_4$
on a particular edge of the oriented octahedron. All the edges of the 
octahedron are equivalent, and $\tau_1$ and $\tau_3$ act simultaneously on all the edges 
according to the same picture.}{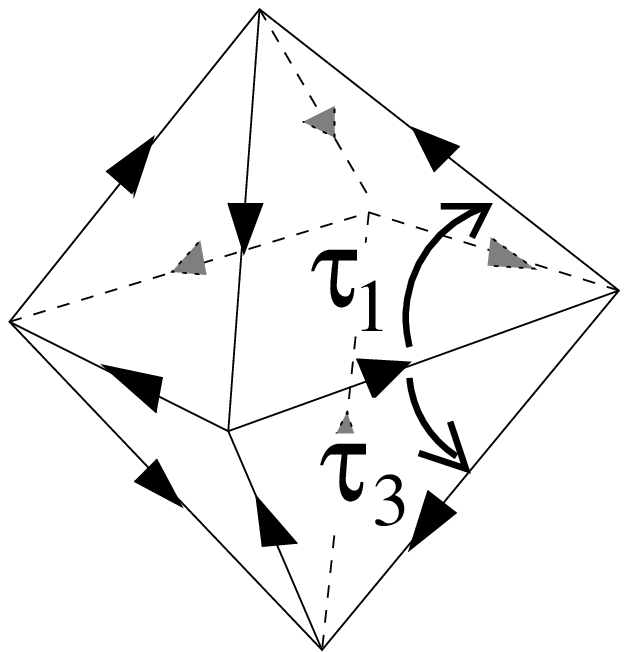}{4.truecm}
\figlabel\elem
\noindent and corresponding to the two elementary mappings shown in Fig.\elem. 
Here we use the standard notation for permutations, by listing their cycles not reduced
to one element (e.g. $(243)$ is the permutation $(1234) \to (1423)$, whereas $(12)(34)$
is the permutation $(1234) \to (2143)$).
	
\noindent In terms of these generators the 12 elements of $A_4$ are:
\eqn\afour{\eqalign{ e ~&=~ \tau_1^3 ~=~ \tau_3^3 ~=~ {\rm identity} \cr
\tau_1 ~&=~ (243) \cr
\tau_3 ~&=~ (134) \cr
\tau_2 \equiv \tau_1^2 ~&=~ (234) \cr
\tau_4 \equiv \tau_3^2 ~&=~  (143) \cr
\tau_1 \tau_3 ~&=~ (132) \cr
\tau_3 \tau_1 ~&=~ (124) \cr
\tau_1^2 \tau_3^2 ~&=~ (142) \cr
\tau_3^2 \tau_1^2 ~&=~ (123) \cr
g \equiv \tau_1 \tau_3^2 ~&=~ \tau_3 \tau_1^2 ~=~ (14)(23) \cr
d \equiv \tau_1^2 \tau_3 ~&=~ \tau_3^2 \tau_1 ~=~ (13)(24) \cr
f \equiv \tau_1 \tau_3 \tau_1  ~&=~ \tau_3 \tau_1 \tau_3 
~=~ (12)(34). \cr }}

\noindent Any other sequence of $\tau_1$ and $\tau_3$ can be reduced to one of the above
elements by use of the four relations\foot{ These relations also imply 
that $\tau_1 \tau_3^2=\tau_3 \tau_1^2$.}
\eqn\relations{\eqalign{\tau_1^3 ~&=~ \tau_3^3 ~=~ {\rm identity} \cr
\tau_1^2 \tau_3 ~&=~ \tau_3^2 \tau_1 \cr
\tau_1 \tau_3 \tau_1 ~&=~ \tau_3 \tau_1 \tau_3 \ .}}
These relations can be easily understood graphically by following the successive
images of a link around the octahedron (see Fig.\elem\ above).
  
\subsec{Vertex rules}

Starting from the link, say $(i,j)$, the four choices of subsequent link \fourchoice\
correspond to the application of the four operators $\tau_1$, 
$\tau_2$, $\tau_3$ or $\tau_4$,
with the correspondence given in Table II below.
%
$$\vbox{\offinterlineskip
\halign{\tv\quad # & \quad\tv \quad
# & \quad \tv \quad  # & \quad \tv #\cr
\noalign{\hrule}
\tvi $z$ & $\sigma$ & group element &\cr
\noalign{\hrule}
\tvi  $\phantom{-}1$ &  $\phantom{-}1$   & \ \ $\tau_1$  &\cr
\tvi  $\phantom{-}1$ &  $-1$  & \ \ $\tau_2$  &\cr
\tvi  $-1$ &  $\phantom{-}1$  & \ \ $\tau_3$  &\cr
\tvi  $-1$ &  $-1$ & \ \ $\tau_4$  &\cr
\noalign{\hrule} }} $$
\par\begingroup\parindent=0pt\leftskip=1cm\rightskip=1cm\parindent=0pt
\baselineskip=11pt
{\bf Table II:}
The correspondence between the $z$ and $\sigma$ variables
and the elements $\tau$ of $A_4$. 
\par
\endgroup\par
%

\noindent A folding of an elementary hexagon of the regular triangular 
lattice corresponds to a product of six basic group elements
$\tau_{\alpha}, \  \alpha \in \{1,2,3,4\}$, chosen from the four
operators above, and such that 
\eqn\product{P\equiv\tau_{\alpha_6}\tau_{\alpha_5}\tau_{\alpha_4}
\tau_{\alpha_3}\tau_{\alpha_2}\tau_{\alpha_1} ~=~ e \ .}

\fig{The six $z_i$ and $\sigma_i$ variables around a given vertex, and the colours $c_i$
of the interior links.}{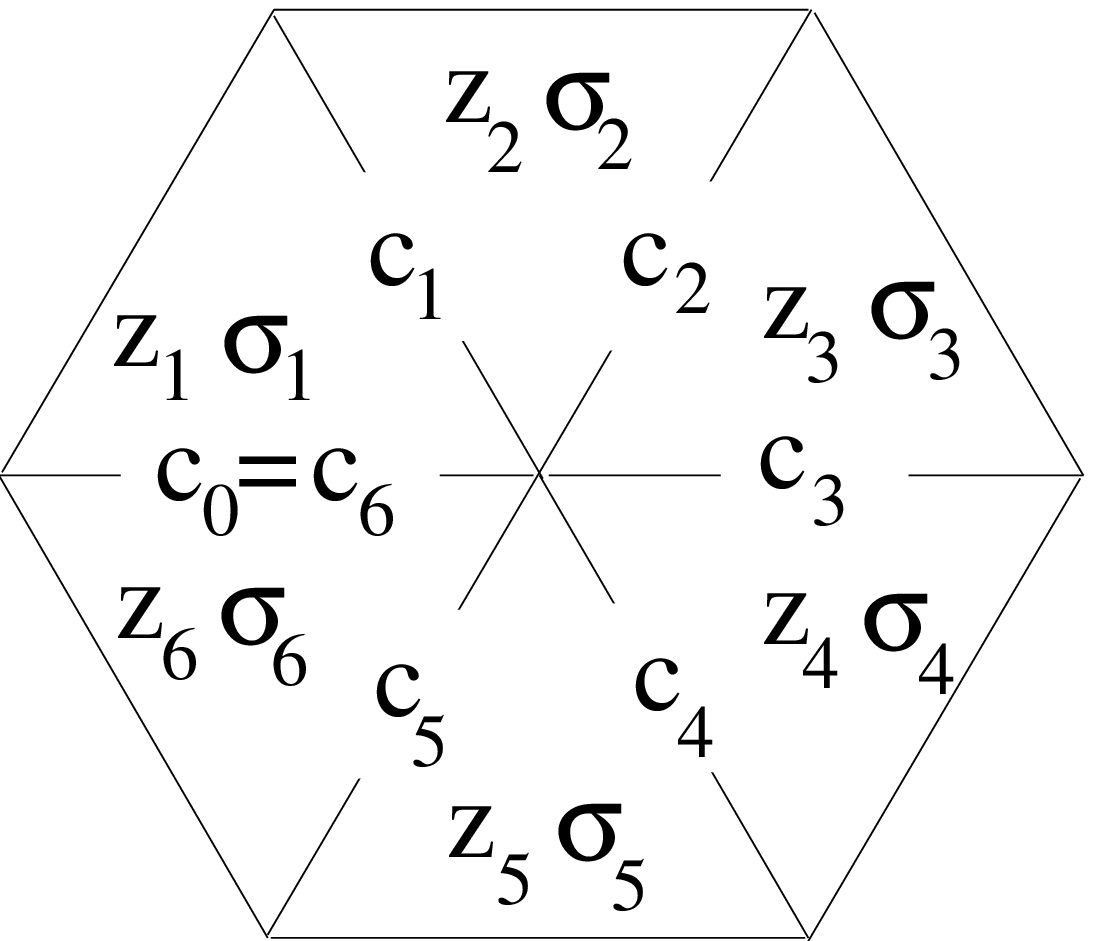}{4.truecm}
\figlabel\spifig
Let us now translate this constraint into folding rules on the six $z$ and $\sigma$ variables
around each vertex (see Fig.\spifig). A first folding rule involves the $\sigma$ variable
alone.  It ensures the 3--colourability of the links of the triangular lattice
in the following way.
Let us assign one of three colours (0=blue ($B$), 1=white ($W$) or 2=red ($R$)) 
to each link of the octahedron.
Choosing the colour of link $(1342)$ to be say, 0 (blue), the colour of a link
$(ijkl)$ is obtained by counting the total number of $\tau_1$ and $\tau_3$ 
operators required to reach this link from $(1342)$. The colour is this
number modulo 3. Since the use of relations \relations\ preserves the total
number of $\tau_1$ plus $\tau_3$ mod 3, the colour is well-defined.
\fig{The assignment of colours for the edges of the octahedron. The four edges
of a given colour lie in the same plane. The three colours around a face are
distinct.}{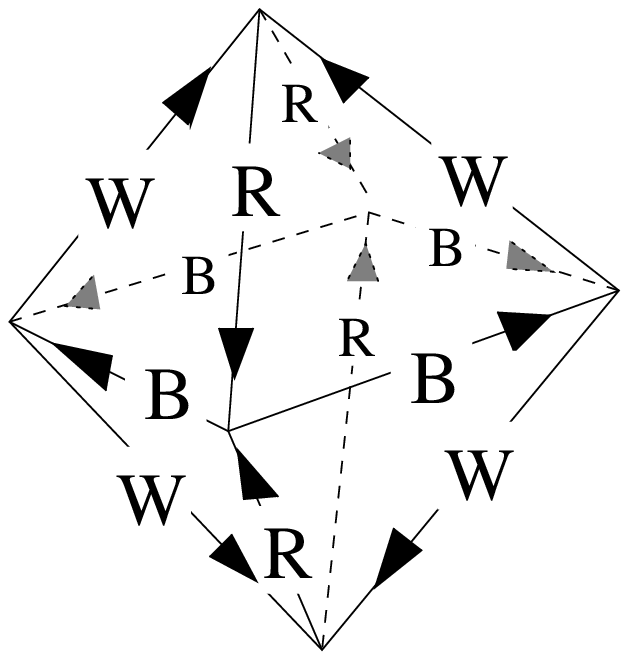}{4.truecm}
\figlabel\colo
\noindent This is illustrated in Fig.\colo: there are 4 blue (resp. white, red)
edges lying in the same plane; the three edges have distinct colour around each face.
Each folding of the triangular lattice induces therefore a 3--colouring of its links.
Conversely, a given 3--colouring of the triangular lattice does not specify its
$3d$ folding state entirely since the same colour may correspond to 4 distinct edges
of the octahedron.

The 3--colourability requirement still leads to a first constraint on the  $\sigma$ variables
around each vertex: the total number of $\tau_1$ and $\tau_3$ mod $3$ in the product $P$
(where we substitute $\tau_2=\tau_1^2$ and $\tau_4=\tau_3^2$) 
must be a multiple of $3$. 
Now in the assignment of colours $\tau_1$ and $\tau_3$ count for $1$, while 
$\tau_2=\tau_1^2$ and $\tau_4=\tau_3^2$ count for $2\equiv-1$ mod $3$. Since this
corresponds precisely to their respective values of $\sigma$,
we find the {\it first folding rule}:
\eqn\ffr{\sum_{i=1}^6 \sigma_i = 0\ {\rm mod}\ 3}
for the six spins around the central vertex of the hexagon.
As mentioned before, this rule first emerged in the 2d folding problem in \DIG, where
it also guaranteed the 3--colourability of the triangular lattice.

In contrast with the $2d$ situation, the restriction \ffr\ is not the only constraint here.
It only ensures that the product $P$ is one of the four colour--preserving elements 
$e,d,g$ or $f$ of \afour.
On the other hand, any sequence of $\tau_1$ and $\tau_3$ satisfying \ffr\ may be 
naturally written
as a product of $e,d,g$ and $f$ operators by simply regrouping the $\tau$'s
into triplets. 
It can be checked that $e,d,g$ and $f$ form a ${\bf Z}_2 \times {\bf Z}_2$
subgroup of $A_4$ with the representation 
\eqn\ztwo{\eqalign{e ~&=~ \tau_1^3 ~=~ \tau_3^3 ~=~ (1,1)  \cr		  
g ~&=~ \tau_1 \tau_3^2 ~=~ \tau_3 \tau_1^2 ~=~ (\eta,1) \cr
d ~&=~ \tau_1^2 \tau_3 ~=~ \tau_3^2 \tau_1 ~=~ (1,\chi) \cr
f ~&=~ \tau_1 \tau_3 \tau_1 ~=~ \tau_3 \tau_1 \tau_3 ~=~
(\eta,\chi), \cr}}
where $\eta^2=\chi^2=1$. 
For the product $P$ to be the identity $e$, we need the number
of $\eta$ and the number of $\chi$ in this decomposition to be separately even.
Consider the product $P$ written as a sequence of $\tau_1$ and $\tau_3$ 
operators and mark the spaces between $\tau$'s by $0,1$ or $2$ consecutively:
  
\epsfysize=0.8truein
\centerline{\epsfbox{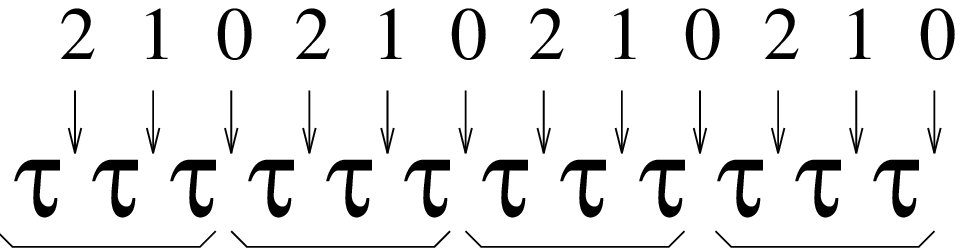}}

The number of $\eta$ (resp. $\chi$) is the number of changes from 
$\tau_1$ to $\tau_3$ or $\tau_3$ to $\tau_1$ occurring at the spaces labelled
by $2$ (resp. $1$).
This follows directly from \ztwo.
On the hexagon these changes can occur {\it only} at the border between two 
neighbouring triangles i.e. on interior links. 
The position of a change ($0,1$ or $2$) is then simply 
the colour of the link at which it occurs.
In terms of $\sigma$ variables, the colour $c_i$ of the internal link 
$i, i=1,...,6$, is given by
\eqn\clink{c_i ~=~ c_0+ \sum_{j=1}^i \sigma_j \ \ {\rm mod} \ 3}
(the first folding rule \ffr\ ensures that $c_6=c_0$).
The quantity ${1- z_i z_{i+1}\over 2}$ is $0$ or $1$ depending
on whether or not a change occurs at link $i$ (with the convention $z_7=z_1$). 
Defining 
\eqn\elfa{\alpha_c \equiv \sum_{i=1}^6 {1-z_i z_{i+1} \over 2} 
\, \delta(c_i , c \ {\rm mod}\ 3)} 
as the total number of changes on the links of colour $c$, $\alpha_c$ must be even.
This gives the {\it second folding rule}
\eqn\sfr{\alpha_c ~=~ 0 \ {\rm mod} \ 2\, : \quad c=1,2 \ .}
Note that \sfr\ implies that $\alpha_0 = 0$ mod 2 since $\alpha_0 + \alpha_1 
+ \alpha_2 = 0$ mod 2, as the total number of changes between $\tau_1$ and $\tau_3$
is even by cyclicity.

To understand this second folding rule, we first
note that a change from $\tau_1$ to $\tau_3$, or vice versa, 
corresponds to a switch to another face of the octahedron, therefore crossing one
of the blue, white or red edge planes bisecting the octahedron (see Fig.\colo). 
The requirement of returning to the {\it same} (say blue) link after $6$ steps
is then equivalent to that of crossing each of the white and red planes 
an even number of times. Indeed $\alpha_1$ (resp. $\alpha_2$) is simply the
number of crossings of the white (resp. red) plane.

The two folding rules \ffr\ and \sfr\ are equivalent to the condition \product\
and therefore characterise the vertex constraint entirely.

\fig{The 96 vertices satisfying the two folding rules \ffr\ and \sfr:
no line corresponds to no fold, a thick line corresponds to a complete fold,
a thin line  corresponds to a fold with obtuse angle and a 
dashed line corresponds to a fold with
acute angle. The degeneracy of each vertex under cyclic
permutations of the links is indicated.}{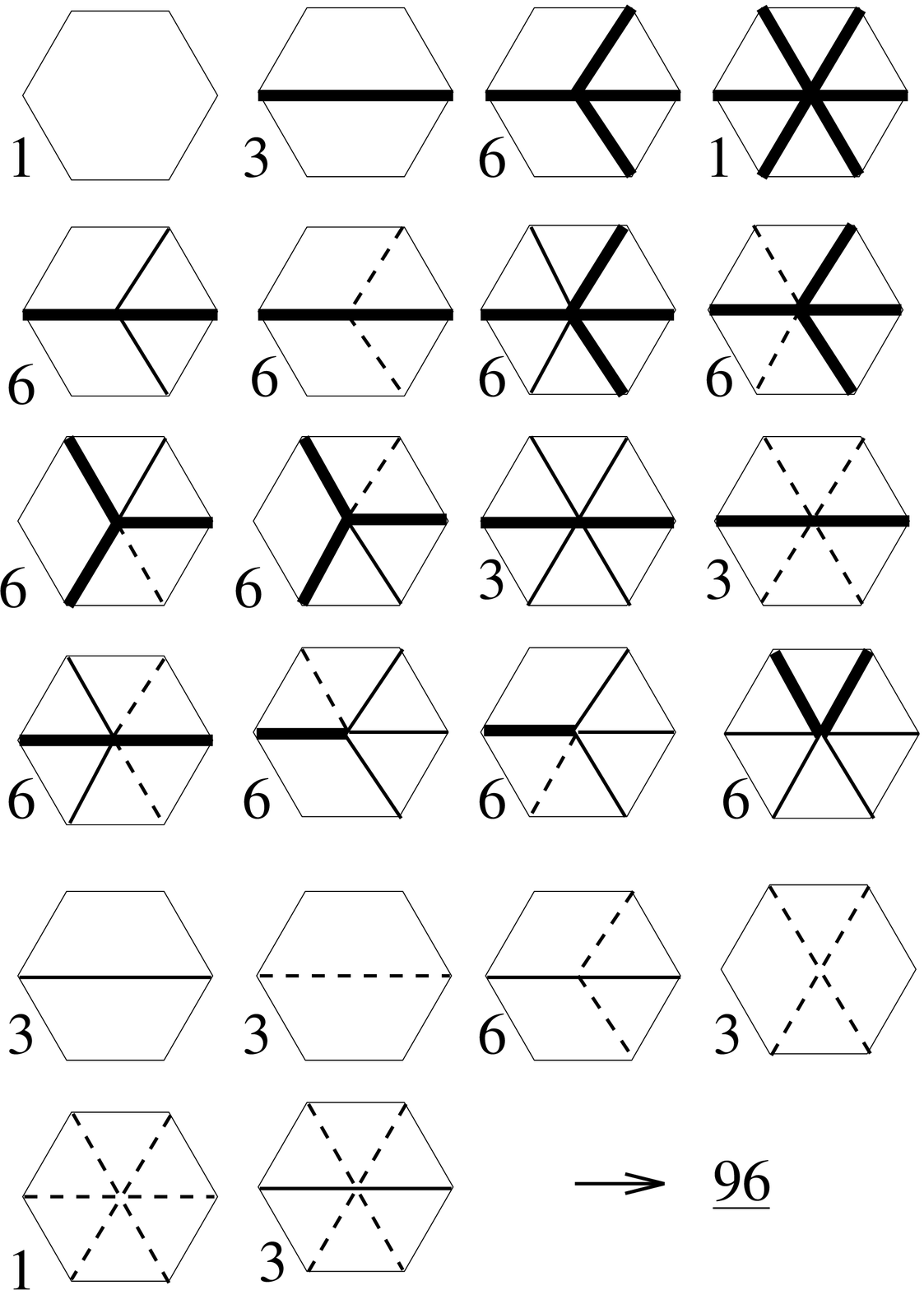}{9.truecm}
\figlabel\ninetysix

With the two folding rules \ffr\ and \sfr\ we find $384=96 \times 4$ vertex
configurations (there is a $4$--fold global degeneracy under reversal of $z$ or $\sigma$), 
as predicted by our previous counting. 
The $96$ folding vertices are displayed in Fig.\ninetysix\ with the following conventions:
no line corresponds to no fold;
a thick line corresponds to a complete folding ($180^{\rm o}$, flip of $\sigma$ only);
a thin line  corresponds to a fold with obtuse angle 
($\arccos(1/3)\sim 70^{\rm o} 32'$ between normal vectors, flip of both
$\sigma$ and $z$) and finally a dashed line corresponds to a fold with
acute angle ($\arccos(-1/3)\sim 109^{\rm o}28'$, flip of $z$ only).
The degeneracy of each vertex under cyclic permutations of the links is also indicated.
\fig{Examples of $3d$ octahedral foldings of an elementary hexagon and the 
corresponding vertices of Fig.\ninetysix.}{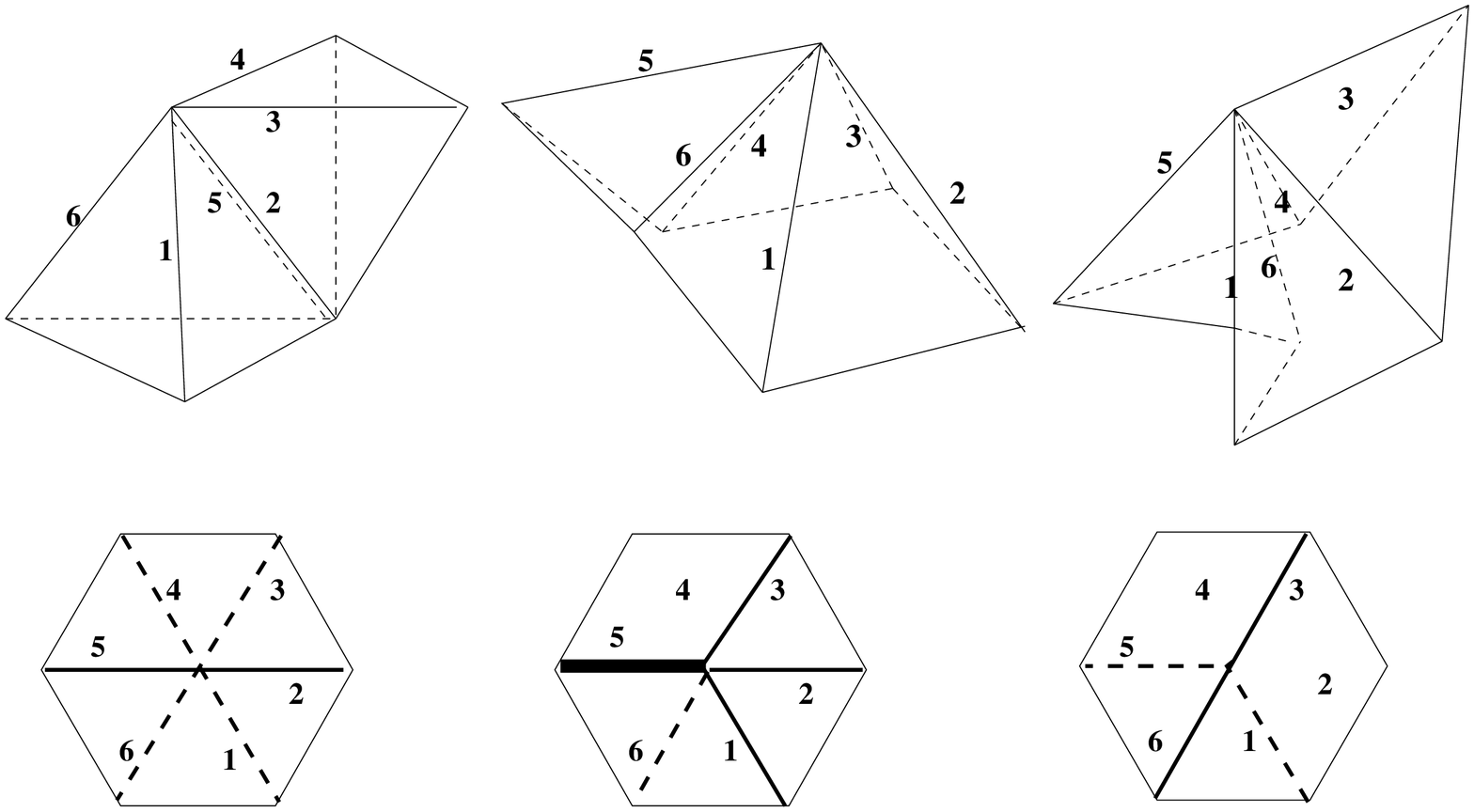}{9.truecm}
\figlabel\exvert
\noindent In Fig.\exvert\ we display a few examples of vertices and the corresponding foldings
in $3d$ space.

The $3d$ octahedral folding problem is now represented as a $96$--vertex model with 
the vertices of Fig.\ninetysix.

\newsec{Entropy: numerical calculation}

In this section, we use the above folding rules to estimate numerically the $3d$ folding
entropy per triangle $s_{3d}={\rm ln} \ q_{3d}$.

The partition function for
a parallelepiped of $2L \times M$ triangles, also conveniently thought of
as a rectangle of $L \times M$ squares divided into two triangles
along one diagonal, can be written
as 
\eqn\partmat{ Z_{L,M}~=~ {\rm Tr} ~ {\cal T}(L)^{M} }
in terms of a row to row transfer matrix ${\cal T}(L)$.

The above reexpression of the folding problem in terms of two face
variables, the spins $z,\ \sigma = \pm 1$, enables us to write 
the transfer matrix from a row of $2L$ triangles to another as a $16^L \times 16^L$
matrix (there are $4$ spin configurations $(z,\sigma)$ on each of the $2L$
triangles). 
We choose for instance to apply free boundary conditions on the right and left
sides of the rectangle.
The entries of the row to row transfer
matrix read
\eqn\transmatelem{\eqalign{ 
{\cal T}(L)_{ \{z,\sigma \},\{z',\sigma' \} }~&=~
{\epsfxsize=7.truecm \epsfbox{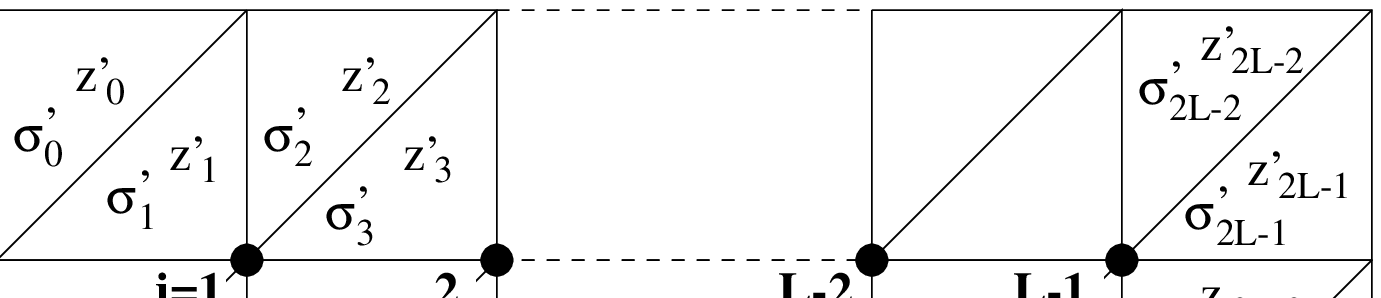}}   \cr
&\ \  \cr
~&=~\prod_{j=1}^{L-1} U_j\times V_j   \cr}}
where $U_j$ and $V_j$ respectively impose the first and second folding rules
\ffr\--\sfr\ around each inner vertex $j$, namely
\eqn\cffr{ 
\eqalign{U_j~&=~ \delta( \sigma_{2j-2}+ \sigma_{2j-1}+ \sigma_{2j}+ 
\sigma_{2j-1}'
+ \sigma_{2j}'+ \sigma_{2j+1}', 0 \ {\rm mod}\ 3) \cr
V_j~&=~ \prod_{c=1}^2  \delta(\alpha_c(z_{2j},z_{2j-1},z_{2j-2},z_{2j-1}',
z_{2j}',z_{2j+1}'),0 \ {\rm mod} \ 2) \cr}}
with $\alpha_c(z_1,z_2,z_3,z_4,z_5,z_6)$ defined by
eq.\elfa.  Note that $V_j$ also depends implicitly
on the $\sigma$'s around  the vertex $j$.

In the large $M$ limit, the partition function \partmat\
is dominated by the largest eigenvalue $\lambda_L$ of the transfer
matrix ${\cal T}(L)$ and the corresponding free energy per row reads
\eqn\domin{ -\lim_{M \to \infty} {1 \over M} {\rm ln} \ Z_{L,M}~=~
- \,{\rm ln} \ \lambda_L  \ .}
The thermodynamic entropy per triangle of the $3d$ folding problem
is therefore
\eqn\frenergy{s_{3d}~=~\lim_{L \to \infty} {1 \over 2L}{\rm ln}\ \lambda_L
~\equiv {\rm ln}\ q_{3d} \ .}

The numerical calculation of $\lambda_L$ is simplified by the fact
that ${\cal T}(L)$ is a sparse matrix, with entries $0$ or $1$.
Actually the total number $N_L$
of non--vanishing elements of ${\cal T}(L)$
reads
\eqn\evanum{ N_L ~=~ \sum_{\{ \sigma,z \},\{ \sigma',z'\}} 
{\cal T}(L)_{ \{z,\sigma \},\{z',\sigma' \} }~=~{\rm Tr} \big({\cal J}[{\cal T}(2)]^L
\big) }
where $\cal J$ is the $256 \times 256$ matrix with all entries equal to $1$ (see \DIGG).
For large $L$, $N_L$ is dominated by the $L$th power
of the largest eigenvalue $\lambda_2$ of 
${\cal T}(2)$, easily computed numerically (see Table III below), 
hence\foot{The largest eigenvalue of ${\cal T}(2)$ matches the value
$14+\sqrt{132}$ up to the precision of our calculation.}
\eqn\lantwo{ N_L ~\propto~ (25.48912...)^L}
to be compared with the total number of elements of ${\cal T}(L)$,
$(256)^L$.

We display the results for the number of non--vanishing elements $N_L$ and 
the largest eigenvalue $\lambda_L$ in Table III below.

%
$$\vbox{\offinterlineskip
\halign{\tv\quad # & \quad\tv \quad
# & \quad \tv \quad  # & \quad \tv \quad  # & \quad \tv #\cr
\noalign{\hrule}
\tvi $L$ & $N_L$& $\lambda_L$ & $(\lambda_L/\lambda_{L-1})^{1 \over 2}$ &\cr
\noalign{\hrule}
\tvi  $1$ &  $256$        & $16.0000000000$  &  $\ $        &\cr
\tvi  $2$ &  $6144$       & $25.4891252930$  & $1.2621688994$ &\cr
\tvi  $3$ &  $153600$     & $44.9037351935$  & $1.3272837214$ &\cr
\tvi  $4$ &  $3891200$    & $83.9628811670$  & $1.3674215892$ &\cr
\tvi  $5$ &  $98992128$   & $162.6428871867$ & $1.3917905107$ &\cr
\tvi  $6$ &  $2521694208$ & $322.3383026700$ & $1.4077917628$ &\cr
\noalign{\hrule} }} $$
\par\begingroup\parindent=0pt\leftskip=1cm\rightskip=1cm\parindent=0pt
\baselineskip=11pt
{\bf Table III:}
The number $N_L$ of non--vanishing transfer
matrix elements and 
the maximum eigenvalue $\lambda_{L}$ for
strips of width $L=1,2,3,...,6$. The ratio $(\lambda_L/\lambda_{L-1})^{1 \over 2}$
is less sensitive to finite size effects than $\lambda_L^{1/2L}$.
\par
\endgroup\par
%

The entries of the matrix ${\cal T}(L)$ are generated by gluing 
together around a central vertex two transfer
matrices of respective sizes $L/2,L/2$ if $L$ is even or $(L-1)/2,(L+1)/2$ 
if $L$ is odd. In this way, the time used for generating ${\cal T}(L)$
becomes comparable to that used for extracting its largest eigenvalue,
by iterative applications of ${\cal T}(L)$ on a vector, which we
normalise at each step (the normalisation factor converges to $\lambda_L$).
The value $L=6$ is obtained by applying some extra symmetry arguments to reduce 
the size of the matrix ${\cal T}(6)$.

Using the Pad\'e--Shanks transformation [\xref\SHAN,\xref\BREZ], we extrapolate
the values of $(\lambda_{L}/ \lambda_{L-1})^{1/2}$ of Table III above
to get a numerical estimate of the partition function per triangle 
\frenergy. We find
\eqn\estipftr{ q_{3d}~\sim~ 1.43(1) \ . }

\newsec{Dressed 3--colouring and various bounds on the entropy}

In this section, we derive exact bounds for the partition function per triangle
$q_{3d}$.
For this purpose, we turn back to the initial definition in terms of tangent
vector link variables.

\subsec{3d folding as dressed 3--colouring}

The link variable of the $3d$ folding problem takes its values among the
$12$ tangent vectors to the octahedron of Fig.\octaor. 
Let ${\bf e}_1,{\bf e}_2,{\bf e}_3$ denote the canonical basis of ${\bf R}^3$ and  
let us fix the positions of the $6$
vertices of the octahedron to be $(\pm 1/\sqrt{2},0,0)$, $(0,\pm 1/\sqrt{2},0)$ and
$(0,0,\pm 1/\sqrt{2})$ in this basis. 
For instance, the face with normal vector ${\bf n}_1$ pointing out in Fig.\octaor\
has the vertices $(1/\sqrt{2},0,0)$, $(0,1/\sqrt{2},0)$, $(0,0,1/\sqrt{2})$.
Denoting by ${\bf x}_i={\bf e}_i/\sqrt{2}$, 
$i=1,2,3$, we list in Table IV below
the $12$ unit tangent vectors grouped according to their outward oriented face label. 
We also display the colour of each vector, as defined in Fig.\colo.
In the present language, the colour of a tangent vector simply corresponds to 
the index of the missing basis vector ${\bf x}$ modulo 3, i.e. $3\equiv 0 \rightarrow B$,
$1 \rightarrow W$ and $2 \rightarrow R$.
%
$$\vbox{\offinterlineskip
\halign{\tv\quad # & \quad\tv \quad
# & \quad \tv \quad  # & \quad \tv \quad # & \quad \tv # \cr
\noalign{\hrule}
\tvi face & normal vector & tangent & colour &\cr
\noalign{\hrule}
\tvi      &   					  &      
$-{\bf x}_1+{\bf x}_2$  & B &\cr
\tvi  $1$ &   ${\bf n}_1=\sqrt{2 \over 3}({\bf x}_1+{\bf x}_2+{\bf x}_3)$ &      
$-{\bf x}_2+{\bf x}_3$  & W &\cr
\tvi      &                                       &      
$-{\bf x}_3+{\bf x}_1$  & R &\cr
\noalign{\hrule}
\tvi  	  &    				       	  &      
$-{\bf x}_1-{\bf x}_2$ & B &\cr
\tvi  $2$ &   ${\bf n}_2=\sqrt{2 \over 3}({\bf x}_1-{\bf x}_2-{\bf x}_3)$ &      
$\phantom{-}{\bf x}_2-{\bf x}_3$  & W &\cr
\tvi      &                                       &      
$\phantom{-}{\bf x}_3+{\bf x}_1$  & R &\cr
\noalign{\hrule}
\tvi      &   					  &      
$\phantom{-}{\bf x}_1-{\bf x}_2$  & B &\cr
\tvi  $3$ &  ${\bf n}_3=\sqrt{2 \over 3}(-{\bf x}_1-{\bf x}_2+{\bf x}_3)$ &      
$\phantom{-}{\bf x}_2+{\bf x}_3$  & W &\cr
\tvi      &                                       &      
$-{\bf x}_3-{\bf x}_1$ & R &\cr
\noalign{\hrule}
\tvi      &  					  &      
$\phantom{-}{\bf x}_1+{\bf x}_2$  & B &\cr
\tvi  $4$ &  ${\bf n}_4=\sqrt{2 \over 3}(-{\bf x}_1+{\bf x}_2-{\bf x}_3)$ &      
$-{\bf x}_2-{\bf x}_3$ & W &\cr
\tvi      &                                       &      
$\phantom{-}{\bf x}_3-{\bf x}_1$  & R &\cr
\noalign{\hrule} }} $$
\par\begingroup\parindent=0pt\leftskip=1cm\rightskip=1cm\parindent=0pt
\baselineskip=11pt
{\bf Table IV:}
The twelve unit tangent vectors around the faces
of the octahedron of Fig.\octaor. We display the face label, the normal
vector to the face (pointing outwards), the three tangent vectors
around the face and their respective colour.
\par
\endgroup\par
%
%

\fig{A typical configuration of tangent vectors around a triangle, expressed in the
${\bf x}$ basis. Once the colour of each link is chosen, namely the three couples 
$({\bf x}_i,{\bf x}_j)$, $({\bf x}_j,{\bf x}_k)$ and $({\bf x}_k,{\bf x}_i)$, 
the three signs $\epsilon_i$, 
$\epsilon_j$ and $\epsilon_k$ are still arbitrary.}{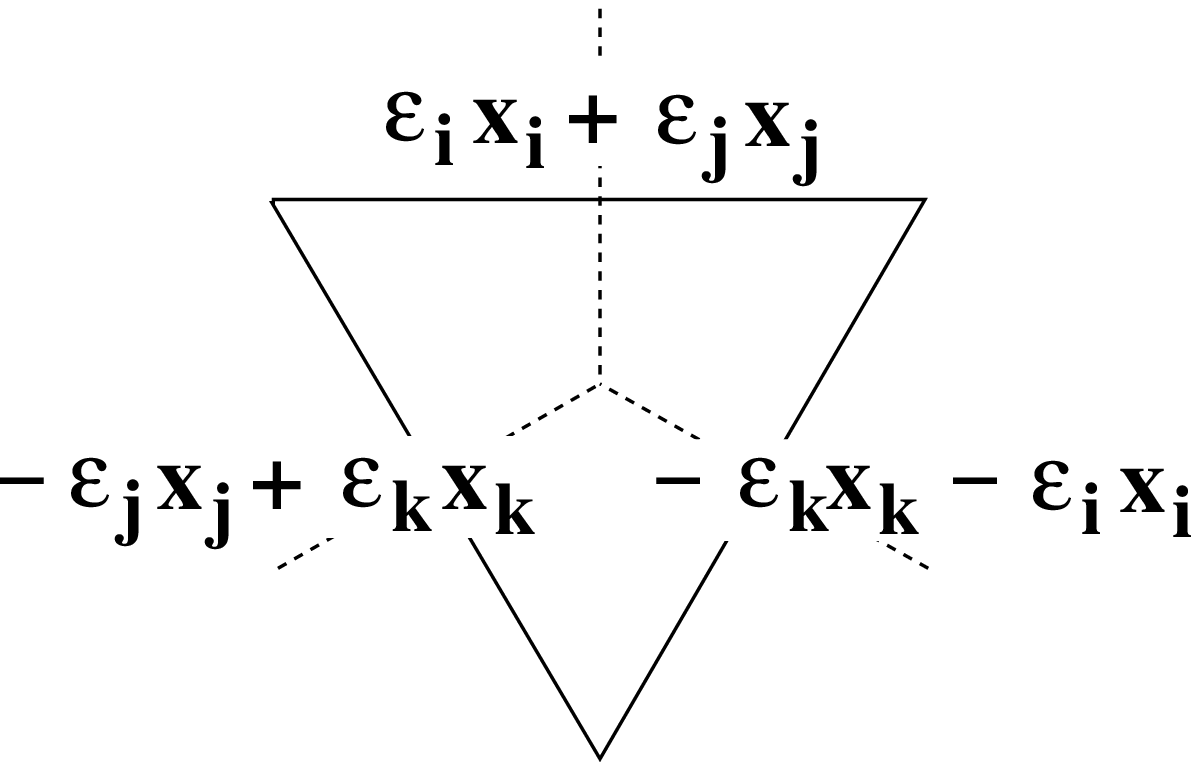}{4.5truecm}
\figlabel\signs

The $48$ configurations of tangent vectors around a face of the triangular 
lattice are specified by the following two sets of data, displayed in Fig.\signs:

(i) one of the 6 allowed colourings of the three links, corresponding to a permutation
of $B,W,R$. This specifies for each link the plane ${\bf x}_i,{\bf x}_j$ where the tangent 
vector lies.

(ii) the assignment of the signs $\epsilon_i,\epsilon_j=\pm$ in the expression 
of the tangent vector to each link: $\epsilon_i {\bf x}_i+\epsilon_j {\bf x}_j$. Due to the
folding rule \foldrule, each ${\bf x}_i$ must appear with the two signs, therefore only three
signs have to be specified for each triangle, leading to $2^3=8$ possibilities.

We recover in this way the $48=6 \times 8$ face configurations mentioned above.
This presentation has the advantage of decoupling the colouring step (i) from the
assignment of signs (ii), therefore displaying the degeneracy of the $3d$ folding
when compared to 3--colouring.
  
For a given 3--colouring of the links of the triangular 
lattice, let us now evaluate the number of choices of signs we can make. 
This is best done on the dual hexagonal lattice, 
whose vertices are the centers of the faces of the triangular lattice, and
whose links (represented by dotted lines in Fig.\signs) cross those of the triangular 
lattice. 
The tangent vectors
are now link variables of the hexagonal lattice. 
A 3--colouring of the links of the triangular lattice
is simply a colouring of the links of the hexagonal lattice, with the constraint
that the three links meeting at each vertex have distinct colours.
Now a choice of sign, say $\epsilon_1$ for the coefficient of
${\bf x}_1$ has to be made on all $B$ and $R$ links. Moreover, this sign propagates
from a given $B$ (resp. $R$) link to the two neighbouring $R$ (resp. $B$) links, 
in order for \foldrule\ to be satisfied at each vertex.
\fig{A sample $3$--colouring of the hexagonal lattice: the $BRBR...$ sequences of links
form dense loops, represented by thick lines. 
The coefficient of ${\bf x}_1$ in the corresponding tangent vectors can be independently fixed 
on each loop to be either $+1$ on the $B$ links and $-1$ on the $R$ links, or
$-1$ on the $B$ links and $+1$ on the $R$ links.}{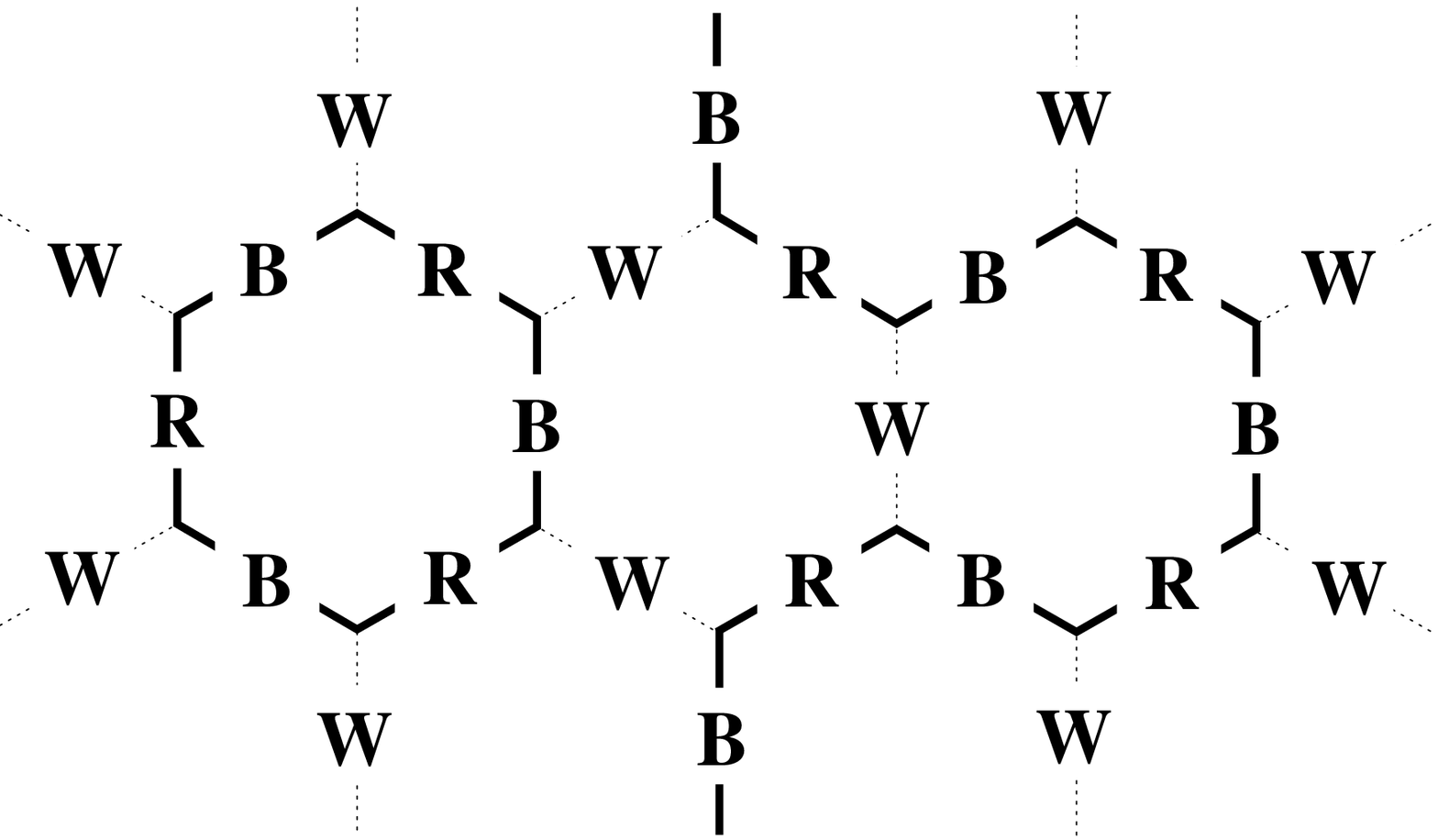}{8.truecm}
\figlabel\bwrloops 
\noindent More precisely, as shown in Fig.\bwrloops, the sequences of $B$, $R$, $B$, $R$, ...
links form closed loops on the hexagonal lattice, and the sign of the ${\bf x}_1$ component
must alternate along each such loop. 
For each $BR$ loop, there are two choices of signs of the ${\bf x}_1$ components:
$\epsilon_1=+$ (resp. $-$) on all $B$ (resp. $R$) links of the loop, or conversely.
The same holds independently
for the choices of signs of the ${\bf x}_2$ components along the 
$BW$ loops and of the ${\bf x}_3$ components along the $WR$
loops.

So for a given colouring of the links of the hexagonal lattice, we
are left with $2$ choices of signs for each of the $m_1$ $BR$ loops,
$m_2$ $BW$ loops and $m_3$ $WR$ loops.
The partition function for $3d$ folding reads then
\eqn\parfuncol{ Z_{3d}~=~ \sum_{3-{\rm colourings}} ~ 2^{m_1+m_2+m_3} }
where the sum extends over all the $3$--colourings of the links of 
the hexagonal lattice and $m_1$ (resp. $m_2$, $m_3$) denote the number
of $BR$ (resp. $BW$, $WR$) loops for each colouring.
The expression \parfuncol\ identifies the $3d$ folding problem of the triangular lattice
with a dressed $3$--colouring problem of the hexagonal lattice, obtained by attaching
a ${\bf Z}_2$ variable $\epsilon$ to each loop of any alternating two colours.
Notice finally that the bi--coloured loops above form three {\it dense} coverings of the hexagonal
lattice, in the sense that each vertex of the hexagonal lattice belongs exactly to one
$BR$, one $BW$ and one $WR$ loop.
The loops of a given type are {\it non--intersecting}.

As an exercise let us re--derive the number $96$ of vertex configurations in this
dressed 3--colouring framework.  This number is simply the partition function of a single 
hexagon $H$ with external legs (dual to an elementary hexagon
of the triangular lattice)
\eqn\numhex{ Z_{H}~=~ \sum_{3-{\rm colourings}\ {\rm of}\ H}~2^{m_1+m_2+m_3}  \ .}
Two situations may occur: 

\noindent (i) the colouring contains a bi--coloured central loop, in which case
one of the $m$'s is equal to $1$ and the two others are equal to $3$ (open loops), 
leading to a weight
$2^7$. There are $6$ such colourings.

\noindent (ii) all the bi--coloured loops are open. Each external leg belongs to two loops and
each loop contains two external legs; the total number of loops is thus equal to the
number of external legs $6$: this leads to a weight $2^6$. There are $6 \times 11 -6$
such colourings (since the total number of 3--colourings of $H$ is $6 \times 11$).

\noindent We finally get 
\eqn\nummm{ Z_{H}~=~6 \times 2^7 + 60 \times 2^6 ~=~ 4608 }
in agreement with \vert.

\subsec{Simple bounds on the entropy}

The simplest lower bound on $q_{3d}=\lim_{N_\Delta \to \infty} Z_{3d}^{1/N_\Delta}$, 
is obtained by minoration of
the sum \parfuncol\ by picking a particular 
$3$--colouring of the hexagonal lattice and evaluating its contribution.

\fig{The antiferromagnetic groundstate which maximises $m_1$, $m_2$ and 
$m_3$ simultaneously. The $BR$ (resp. $BW$, $WR$) hexagonal loops are indicated
by discs of different colours.}{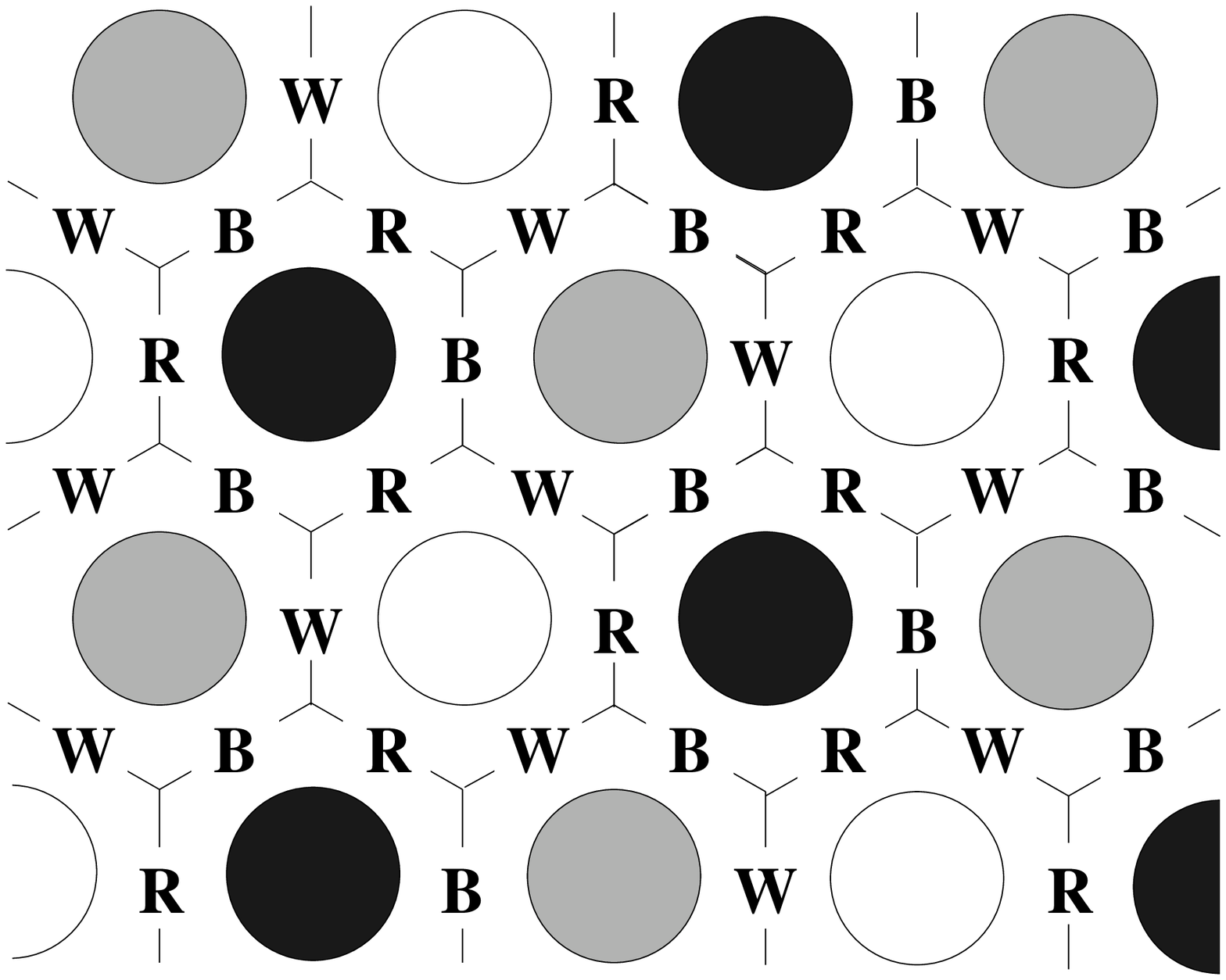}{6.truecm}
\figlabel\fundamental
We take the particular configuration depicted in Fig.\fundamental, 
which maximises $m_1$, $m_2$ and $m_3$ simultaneously.
This configuration corresponds to a regular antiferromagnetic
ordering of the $\sigma$ variables. The $B$ and $R$ links are arranged in a dense
regular set of hexagonal loops, as well as the $B$ and $W$ links and the $W$ and $R$ links.
Since the smallest loop on the hexagonal lattice has length 6, 
$m_1$, $m_2$ and $m_3$ are clearly maximal.
Noticing that the total number $N_H$ of hexagons in the hexagonal lattice is 
equal to the total number $N_V=N_\Delta/2$ of vertices of the dual triangular lattice, 
this configuration has
\eqn\compfold{ m_1~=~m_2~=~m_3~=~{N_H \over 3}~=~{N_V \over 3}~=~{N_\Delta \over 6} \ . }
Hence $Z_{3d} > 2^{N_\Delta /2}$, so that
\eqn\debile{ q_{3d}~\geq ~ \sqrt{2}~=~1.414... }
In the 96--vertex language, the contribution of the antiferromagnetic
groundstate of Fig.\fundamental\ corresponds to a restriction of the model
to the subset of vertices 
with alternating $\sigma$ variables.  
This subset is characterized by the presence of a
thick or a thin solid line on {\it each} internal link (see Fig.\ninetysix). There
are $1+6+3+6=16$ such vertices. In this respect, $\sqrt{2}$ is simply
the exact partition function per triangle 
of this restricted $16$--vertex model, as can be checked directly.

We can improve this bound by incorporating more configurations in our counting, 
i.e. excitations of the antiferromagnetic groundstate of Fig.\fundamental. 
This is done in the Appendix, leading to the improved lower bound
\eqn\implobo{ q_{3d} ~\geq ~ 1.429... }

Using now the fact that, for an arbitrary configuration of 3--colouring of the hexagonal lattice,
the numbers $m_1$, $m_2$ and $m_3$ are always smaller or equal to $N_{\Delta}/6$, we also get
a simple upper bound
\eqn\upperbou{ Z_{3d} ~\leq ~ 2^{N_\Delta /2} ~\sum_{3-{\rm colourings}}~ 1 ~ =
~2^{N_\Delta /2}~Z_{2d} }
since the $2d$ folding problem and that of 3--colouring of the hexagonal lattice are
equivalent \DIG. Thus, we get an upper bound on $q_{3d}$ in terms of $q_{2d}$ \exatwod\
\eqn\uppercut{ q_{3d} ~\leq ~ \sqrt{2} ~ q_{2d}~=~1.709...}

\subsec{Improved bounds from $2d$ folding in a field}

The partition function of the $2d$ folding problem also has an expansion in terms of {\it dense}
loops on the hexagonal lattice.  As mentioned above, the $2d$ folding problem is equivalent
to that of $3$--colouring of the links of the hexagonal lattice, dual to the original triangular
lattice.  Instead of considering the $3$--colourings of the hexagonal lattice links, however, 
we can concentrate on say the $W$--coloured links only. 
Consider a particular $3$--colouring of the links. We have seen above how the paths
of $BRBR...$ links  form dense, non--intersecting closed loops on the hexagonal lattice. 
Exchanging the $B$ and $R$ links along any of 
these loops independently leads to equally admissible $3$--colourings of the links.
Therefore, for a given {\it admissible} configuration of $W$ links, one is left with $2$ possible
independent choices of colourings per $BR$ loop. 
The number $m_1$ of such loops is fixed
by the position of the $W$ links only. We can write
\eqn\twodback{ 
Z_{2d}~=~ \sum_{3-{\rm colourings}}~ 1~=~ \sum_{W \ {\rm link}\ {\rm config.}}~
2^{m_1}  \ .}
This is the $2d$ version of the $3d$ folding dense loop expression \parfuncol.

The solution by Baxter \BAX\ of the $3$--colouring problem of the links of the hexagonal lattice
includes the introduction of an extra parameter in the weighting of
configurations. 
This parameter may be interpreted as a {\it staggered magnetic field}
$h_{st}$ in the following way.  
In the language of the
face spin ($\sigma$) variable ($\sigma=+1$ or $-1$ on a face indicates whether the colours
of adjacent links around the face increase ($B \to W \to R \to B$) or decrease ($B \to R \to W \to B$)
counterclockwise),  this weight reads
\eqn\weibax{\eqalign{ e^{\sigma h_{st}} &\quad \hbox{ per triangle facing up,} \ \Delta \ {\rm,}\cr
e^{-\sigma h_{st}} &\quad \hbox{ per triangle facing down,} 
\ \nabla  \ {\rm.}\cr}}
This gives a maximal weight to the completely folded (antiferromagnetic) groundstate of the 
model (see Fig.\fundamental). Baxter's result is the exact partition function per triangle
\eqn\baxres{ q_{2d}(h_{st})~=~ z ~\prod_{n=1}^\infty { (1 - z^{12(1-3n)}) \over \sqrt{
(1-z^{12(2-3n)})(1-z^{12(-3n)})} } }
with $z=e^{h_{st}}$. The abovementioned result \exatwod\ for the $2d$ folding entropy is 
then recovered in the limit $h_{st}\to 0$, i.e. $z \to 1$.

\fig{The four basic possibilities of local orientation for the succession of a $B$
and a $R$ oriented link on the hexagonal lattice (up to global rotation by $\pm 2 \pi /3$). 
The corresponding state of the visited (dual) triangle 
is represented above each two--link state. The weight in a staggered magnetic field is 
$e^{h_{st}}$ in the first and third case, and $e^{-h_{st}}$ in the second and fourth ones. 
The former correspond to left turns, whereas the latter correspond to right turns along
the $BR$ loop.
Exchanging the $B$ and $R$ links would simply reverse the sign of the face spin $\sigma$
as well as the orientation of the links.}{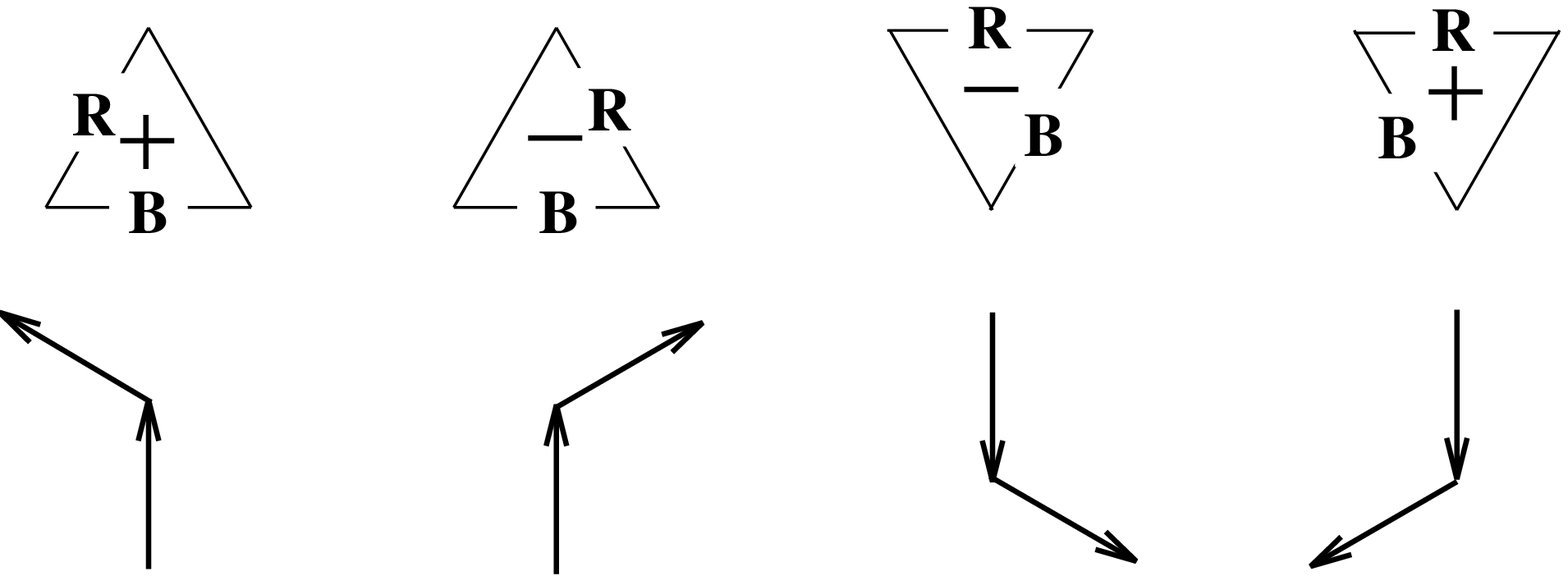}{7.truecm}
\figlabel\twopos
Let us now translate the result \baxres\ into the language of dense $BR$ loops \twodback\
on the hexagonal lattice.  Consider a particular admissible
configuration of $W$ links (a term in the sum in \twodback).
The extra weights \weibax\ can be translated into a dressing of the corresponding 
$BR$ loops.  More precisely, a succession of two $B$ and $R$ links along a $BR$ loop
may be in one of the $4$ states (up to a rotation by $\pm 2 \pi /3$) 
depicted in Fig.\twopos. This provides us with a natural orientation of the loops,
according to Fig.\twopos. Each link joins the centers of two adjacent triangles, one pointing up
and the other pointing down. 
We orient the $B$ links toward the upward pointing triangle, and
the $R$ links toward the downward pointing triangle. Since 
triangles pointing up and down alternate along a loop, as well as $B$ and $R$ links, this 
rule defines a consistent orientation for each $BR$ loop. 
Reversing the
orientation of a loop amounts to exchanging its $B$ and $R$ links, with fixed $W$ links.
By inspection of the $4$ cases of Fig.\twopos, 
we see that the weights \weibax\ translate into local weights 
\eqn\locwei{\eqalign{
e^{h_{st}}~&\quad \hbox{per left turn,}\cr
e^{-h_{st}}~&\quad \hbox{per right turn} \cr }}
of each oriented $BR$ loop.  
The previous degeneracy of $2$ per loop now amounts to summing over the two orientations 
of each loop, resulting in a total weight
\eqn\weiloop{ e^{6 h_{st}} + e^{-6 h_{st}} ~=~ 2 \cosh 6 h_{st}}
per loop, as a closed loop has its number of left turns minus its number of right turns
equal to $\pm 6$, and the opposite for the opposite orientation.
The partition function for the $2d$ folding problem in a staggered magnetic
field reads therefore
\eqn\stagpart{\eqalign{
Z_{2d}(h_{st})~&=~ \sum_{W \ {\rm link}\ {\rm config.}}~ (2 \cosh 6 h_{st} )^{m_1} \cr
&\propto q_{2d}(h_{st})^{N_{\Delta}} \cr} }
where the sum is as in \twodback\ and $q_{2d}(h_{st})$ is defined in \baxres.
To get a more symmetric form, we can sum as well over the $3$--colourings of the links
of the hexagonal lattice, thus removing a factor $2^{m_1}$ in the above expression, hence
\eqn\altercol{\eqalign{
Z_{2d}(h_{st})~&=~ \sum_{W \ {\rm link}\ {\rm config.}}~ (\cosh 6 h_{st} )^{m_1} 
\times 2^{m_1} \cr
&=~ \sum_{W \ {\rm link}\ {\rm config.}}~ (\cosh 6 h_{st} )^{m_1} \sum_{B, R} ~1 ~\cr
&=~\sum_{3-{\rm colourings}}~ (\cosh 6 h_{st} )^{m_1}  \ . \cr}}
In the last expression, we could replace $m_1$ by $m_2$ or $m_3$ without distinction.

With this function at hand, we are now ready to give more bounds on $q_{3d}$.
Starting from a given $3$--colouring configuration of
the dense loop expression \parfuncol, let us examine the numbers
$m_1$, $m_2$ and $m_3$ in more detail. A given $BR$ loop is called {\it direct}
(resp. {\it indirect}) iff its total weight \locwei\
is $e^{6 h_{st}}$ (resp. $e^{-6 h_{st}}$).
We can therefore decompose the number $m_1$ of $BR$ loops into
the numbers $d_1$ and $i_1$ of direct and indirect $BR$ loops respectively. Analogously,
we can define the numbers $d_2$, $d_3$, $i_2$, $i_3$ of direct and indirect $BW$ and $WR$ loops.
We have the following relations
\eqn\monedir{\eqalign{
m_j~&=~ d_j+ i_j \ , \quad j=1,2,3 \cr
d_1-i_1~&=~ d_2-i_2~=~ d_3-i_3 \ .\cr}}
To prove the last identity,  we note that for a fixed 3--colouring, the total weight \weibax\
can be evaluated in terms of any of the three systems of dense loops ($BR$, $BW$ or $WR$)
independently, with the same result
\eqn\weitotbrw{ e^{6 h_{st} (\sum_\Delta \sigma -\sum_\nabla \sigma)}
~=~e^{6 h_{st} (d_1-i_1)}~=~e^{6 h_{st} (d_2-i_2)}~=~e^{6 h_{st} (d_3-i_3)}\ . }

A first lower bound on $q_{3d}$ can be obtained as follows.
In the dense loop expression \parfuncol\ of the partition function $Z_{3d}$, 
we restrict the sum to the configurations where all the $BR$ loops are direct. 
For each admissible
colouring of the $W$ links, there is exactly one such configuration of $B$ and $R$ links
(it corresponds to a particular choice of orientation on each loop).
By summing over these states, we get a minoration $Z_{BR}^{dir}$ of $Z_{3d}$.
All these states have by definition $i_1=0$, hence $m_1=d_1$. 
Using the identities \monedir, we find that
\eqn\ineg{\eqalign{ m_2~&=~d_2+ i_2~=~(d_2 -i_2)+ 2 i_2 ~=~ 
d_1+ 2 i_2~=~m_1+2 i_2 ~\geq~ m_1 \ ,\cr
m_3~&=~ d_3 + i_3~=~ (d_3 -i_3)+ 2 i_3 ~=~ d_1+2 i_3~=~ m_1 + 2 i_3~ \geq ~ m_1 \ .\cr}}
The partition function
can therefore be bounded from below by
\eqn\belowpart{ Z_{BR}^{dir}~=~\sum_{W \ {\rm link} \ {\rm config.}\atop
{\rm all}\ BR \ {\rm loops}\ {\rm direct}}~ 2^{m_1+m_2+m_3} \geq 
\sum_{W\ {\rm link}\ {\rm config.}}~ 2^{3 m_1}  \ .}
The latter is readily identified as the partition function \stagpart\ 
of $2d$ folding in a staggered magnetic field $h_{st}^*$, such that
\eqn\statwod{ 2 \cosh 6 h_{st}^*~=~2^3 }
i.e. 
\eqn\hachsta{h_{st}^*~=~{1 \over 6} {\rm ln} \ (4+\sqrt{15}) \ .}
Finally we get the bound
\eqn\lowfromtwo{ q_{3d} ~\geq~ q_{BR}^{dir} ~\geq~ q_{2d}(h_{st}^*)~=~1.421... }

If for a given $W$ link configuration, we now take into account {\it all} the orientations
of the $BR$ loops, 
eq.\ineg\ no longer holds, but we still have the inequalities
\eqn\ineqbr{ \eqalign{
m_2~=~ (d_2 -i_2)+ 2 i_2 ~=~(d_1 -i_1)+2i_2~=~m_1 +2i_2 -2 i_1 ~\geq~
m_1 -2i_1 \ , \cr
m_3~=~ (d_3 -i_3)+ 2 i_3 ~=~(d_1 -i_1)+2i_3~=~m_1 +2i_3 -2 i_1 ~\geq~
m_1 -2i_1 \ , \cr}}
so that we can write
\eqn\betmin{ Z_{3d} ~\geq~ \sum_{3 - {\rm colourings}}~ 2^{3 m_1 -4 i_1}~=~
\sum_{W \ {\rm link}\ {\rm config.}}~ 2^{3 m_1}~\sum_{B,R \ {\rm links}} 2^{-4i_1} }
where we simply separate the sum over the admissible $W$ links from
that over the $B$ and $R$ links
along $BR$ loops. Since the $m_1$ $BR$ loops can be chosen to be direct or indirect
independently, 
the latter sum factorises and contributes for
\eqn\facontr{\sum_{i_1=0}^{m_1}~{ i_1 \choose m_1}~{1 \over 2^{4 i_1}}~=~ 
(1 + {1 \over 2^4})^{m_1}~=~ (17/16)^{m_1} \ , }
and hence we finally get
\eqn\lowbest{ Z_{3d}~\geq~ \sum_{W \ {\rm links}}~ (17/2)^{m_1} \ .}
According to eq.\stagpart, the latter partition sum is nothing but that of 
$2d$ folding in staggered magnetic field $h_{st}^{**}$, such that
\eqn\stapro{ 2 \cosh 6 h_{st}^{**}~=~{17 \over 2} }
i.e. 
\eqn\autrh{ h_{st}^{**}~=~ {1 \over 6} {\rm ln}\ {17+\sqrt{273} \over 4} \ . }
This gives the following lower bound on $q_{3d}$
\eqn\lowverybest{ q_{3d}~\geq ~q_{2d}(h_{st}^{**})~=~ 1.4351799...}

Comparing this bound with the numerical estimate \estipftr, we suspect that the 
exact result for $q_{3d}$ differs from this value by no more than $1$ percent.

We can also obtain a majoration of $Z_{3d}$ by use of the H\"older inequality on averages, namely
\eqn\holder{ \langle A \times B \times C \rangle ~\leq ~ 
\langle A^\alpha \rangle^{1 \over \alpha}
\langle B^\beta \rangle^{1 \over \beta} \langle C^\gamma \rangle^{1 \over \gamma} }
where $1/\alpha+1 /\beta + 1/ \gamma=1$. The majoration reads
\eqn\majo{\eqalign{
Z_{3d}~&=~\sum_{3-{\rm colourings}}~ 2^{m_1} \times 2^{m_2} \times 2^{m_3} \cr
&\leq ~ \left( \sum_{3-{\rm colourings}}~ 2^{\alpha m_1} \right)^{1 \over \alpha} \times \cr
&\ \times \left( \sum_{3-{\rm colourings}}~ 2^{\beta m_2} \right)^{1 \over \beta} \times \cr
&\ \times \left( \sum_{3-{\rm colourings}}~ 2^{\gamma m_3} \right)^{1 \over \gamma} .\cr}}
The $3$ terms on the rhs of \majo\ are readily identified as powers of $2d$ folding
partition sums \stagpart\ with respective staggered magnetic fields
\eqn\staplus{
h_{st}(x)~=~ {1 \over 6} {\rm ln} \ (2^{x} + \sqrt{2^{2x}-1}) \quad 
x=\alpha,\beta,\gamma .}
The lowest upper bound provided by \holder\ corresponds in fact
to $\alpha=\beta=\gamma=3$, which leads to
\eqn\upperbound{ q_{3d} ~\leq ~ q_{2d}(h_{st}(3))~=~1.589469...}

\newsec{Discussion}

\subsec{Estimating $q_{3d}$}

In our strategy for getting lower bounds on $q_{3d}$,  we first considered the
contribution of a particular state, the groundstate of Fig.\fundamental. 
In this state, the $W$ links form a regular hexagonal pattern, and the $m_1=N_\Delta /6$ $BR$
loops are all direct.  This leads to the first estimate $\sqrt{2}$ of eq.\debile.
Keeping the $W$ links fixed, elementary excitations are obtained by exchanging the $B$ and $R$
links along some loops, thus reversing their orientation from direct to indirect.
As shown in the Appendix, these excitations have a fugacity $1/2^4$ along with some interactions.
Ignoring the interactions leads to the improved lower bound $\sqrt{2}(1+1/2^4)^{1/6}$ (A.3).
Taking these interactions into account leads to the lower bound 
$q_{BR}$ of (A.5), expressed in terms of the entropy of a loop gas on the hexagonal lattice,
reproducing the high temperature expansion of the $O(n=4)$ model at coupling
$K=1/2$ (A.8).

In parallel, we also considered the sum over all admissible configurations of $W$ links. 
With all $BR$ loops direct, we get the first estimate $q_{2d}(h_{st}^*)$ of eq.\lowfromtwo,
with $2 \cosh 6 h_{st}^*=\sqrt{2}^6$. For each $W$ link configuration, elementary excitations
also correspond to reversing the orientation of some $BR$ loops from direct to indirect.
There also, we were led to assign a weight $1/2^4$ per indirect loop (see \betmin), and
we obtained the bound $q_{2d}(h_{st}^{**})$ of eq.\lowverybest\ with
$2 \cosh 6 h_{st}^{**}=\big(\sqrt{2}(1+1/2^4)^{1/6} \big)^6$.
It is therefore tempting to conjecture the lower bound
\eqn\loconj{ q_{3d} ~\geq ~q_{2d}(h_{st}^{***}) }
with 
\eqn\conjcosh{ 2 \cosh 6 h_{st}^{***}~=~ \big(q_{BR} \big)^6 }
to account for interactions between the indirect loops, with $q_{BR}$ given by (A.8).  
This lower bound could in fact be
the exact value of $q_{3d}$ since it would incorporate all the effects of excitations.
$q_{BR}$ is estimated through a Mayer expansion in eq.(A.4).  With this estimation, we get
\eqn\llooconj{ q_{2d}(h_{st}^{***})~\sim~ 1.4356... }

\subsec{$d$--dimensional generalisation}

The $3d$ octahedral folding problem has a natural generalisation to ${\bf R}^d$, by considering
folding on a generalised $d$--dimensional FCC lattice.
Equivalently, this is done by restricting the
$d$--dimensional tangent vectors to be the edges of oriented polytope of ${\bf R}^d$ 
generalising the octahedron, whose $2 \times d$ vertices have positions
$\pm {\bf e}_i/ \sqrt{2}$, where ${\bf e}_i$, $i=1,...,d$ is the canonical basis of ${\bf R}^d$.
The $2d(d-1)$ unit tangent vectors read $\epsilon_i {\bf x}_i+\epsilon_j {\bf x}_j$ 
(${\bf x}_i={\bf e}_i / \sqrt{2}$ as before). They form $4 d(d-1)(d-2)/3$ 
triplets with vanishing sum, corresponding to the faces of the polytope. This
gives $3! \times 4 d(d-1)(d-2)/3=8d(d-1)(d-2)$ possible environments for a given triangle.
Any such environment still takes the form displayed in Fig.\signs, where now
$1 \leq i \neq j \neq k \leq d$. As in the $3d$ case, let us consider the tangent vectors
as link variables on the dual hexagonal lattice. The tangent vectors
with non--zero ${\bf x}_i$ component form $m_i$ loops along which the sign $\epsilon_i$ alternates,
for $i=1,...,d$.  These loops form a {\it dense} covering of the hexagonal lattice in the following way:
any vertex of the hexagonal lattice belongs to exactly three loops of different type, 
and any link of the hexagonal lattice belongs to exactly two loops of different type.
As in the $3d$ case, the two choices of $\epsilon_i$ signs per loop lead to a
partition function
\eqn\partdd{ Z_{d}~=~ \sum_{{\rm dense}\ {\rm loops}} ~ 2^{m_1+m_2+...+m_d} \ . }

To compute the number of possible vertex environments, we evaluate the partition function
of a single hexagon $H$ with  external legs (as in \numhex)
\eqn\dhex{\eqalign{ Z_{d,H}~&=~ \sum_{{\rm dense} \ {\rm loops} \atop {\rm on} \ H} 
2^{m_1+\cdots+m_d} \cr
&=~ 2^7 r_d + 2^6 s_d \cr}}
where $r_d$ (resp. $s_d$) denotes the number of configurations with (resp. without) an internal closed
loop: the first term corresponds to $7$ loops ($1$ closed internal loop and $6$ open loops
entering and exiting the hexagon through the external legs), whereas the second only
has a total number of $6$ loops.
Let us first compute the number $r_d$ of configurations with a closed internal loop. We
fix the type of the central loop, say to ${\bf x}_1$ (among the $d$ possible choices).
Around each vertex of the hexagon, as the ${\bf x}_1$ loop occupies the two inner (left and right)
links, the 
only possibility is that a loop of type ${\bf x}_i$ occupies the left and external links, 
whereas a loop of type ${\bf x}_j$ occupies the right and external links, with
$2 \leq i \neq j \leq d$. 
This suggests the introduction of a transfer matrix $M^{(0)}$ of 
size $(d-1) \times (d-1)$ mapping an internal link to the subsequent one, with entries
\eqn\dtra{ M^{(0)}_{i,j}~=~ (1 - \delta_{i,j}) \quad 2 \leq i,j \leq d\ .}
We find
\eqn\xdx{ r_d~=~ d \ {\rm Tr} \ \big[ M^{(0)} \big]^6~=~d(d-2)\big( (d-2)^5 + 1\big) }
where the prefactor $d$ accounts for the $d$ possible choices of the type of the internal loop.
Next let us evaluate the total number $r_d+s_d$ of loop configurations 
on $H$. Here again, we need to construct a transfer matrix from an internal link to the 
subsequent one. The state of a link is specified by the two types $i<j$ of loops to
which it belongs, namely $d(d-1)/2$ choices.  The desired transfer matrix has therefore
the size ${ d(d-1) \over 2} \times { d(d-1) \over 2}$, and its entries read
\eqn\trand{ M_{ij;kl}~=~ \delta_{ik} (1 -\delta_{jl}) +
\delta_{jl}(1 - \delta_{ik}) + \delta_{il} + \delta_{jk} }
for $1 \leq i<j \leq d$ and $1 \leq k < l \leq d$. Note that this matrix is slightly different from 
that used for the $3d$ folding problem \connectivity. 
After some algebra, we find
\eqn\resdtr{ r_d + s_d~=~ {\rm Tr}\  M^6~=~d(d-1)(d-2)(d^4+42 d^3-380 d^2+1096 d-1072) \ .}
Using \xdx\ and \dhex, we finally get
\eqn\hexdpar{ Z_{d,H}~=~64 d(d-1)(d-2) (2d^4+33d^3-349 d^2 +1047 d -1041)  \ .}
Factoring out the $8d(d-1)(d-2)$ configurations of a given triangle, 
the $d$--dimensional folding problem is therefore equivalent to a $V_d$--vertex model,
with 
\eqn\vertexd{ V_d~=~8(2d^4+33d^3-349 d^2 +1047 d -1041)  \ .}
We recover $V_3=96$ for the $3d$ folding problem, whereas $V_4=1496$, $V_5=6752$, $V_6=19176$...

\newsec{Conclusion}

In this paper, we have defined the $3d$ octahedral folding i.e. the folding of the triangular lattice
on the $3d$ FCC lattice. This model was formulated as a 96--vertex model, with two face spin variables
$z$ and $\sigma$ subject to the two folding rules \ffr\ \sfr. Equivalently,  the partition
function of this model was reexpressed as that of a dressed 3--colouring problem,  involving
a dense covering of the hexagonal lattice by bi--coloured loops.  With these two 
formulations at hand, we were able to estimate the folding entropy $s_{3d}={\rm ln}\ q_{3d}$,
both numerically by use of a transfer matrix and analytically by deriving various 
exact bounds.

Beyond mere counting of folding states,
it would be interesting to obtain the complete phase diagram of this system,
including both bending rigidity and magnetic field as was performed in \DIGG\ for the
$2d$ case.  
In particular, it would be desirable to know the precise status of the crumpling transition in this
framework, including its order (first or second).

\noindent{\bf Acknowledgements}

The research of M.B. was supported by the Department of Energy, USA, under contract 
N$^{\rm o}$ DE-FG02-85ER40237.
M.B. and E.G. are also grateful for support under  NSF grant N$^{\rm o}$ PHY89-04035
from  the Institute for Theoretical Physics at Santa Barbara, where this work was initiated.
We thank J.-M. Normand for a critical reading of the manuscript.

\appendix{A}{Lower bounds on the entropy from local excitations}

\fig{An elementary $BR$ excitation of the antiferromagnetic groundstate of Fig.\fundamental. 
The $B$ and $R$ links are exchanged on the shaded hexagon. 
The three neighbouring $BW$ hexagons are glued to form 
one loop (thick solid line on the first figure), whereas similarly the three
neighbouring $WR$ hexagons are glued to form one loop (thick dashed lines 
on the second figure).}{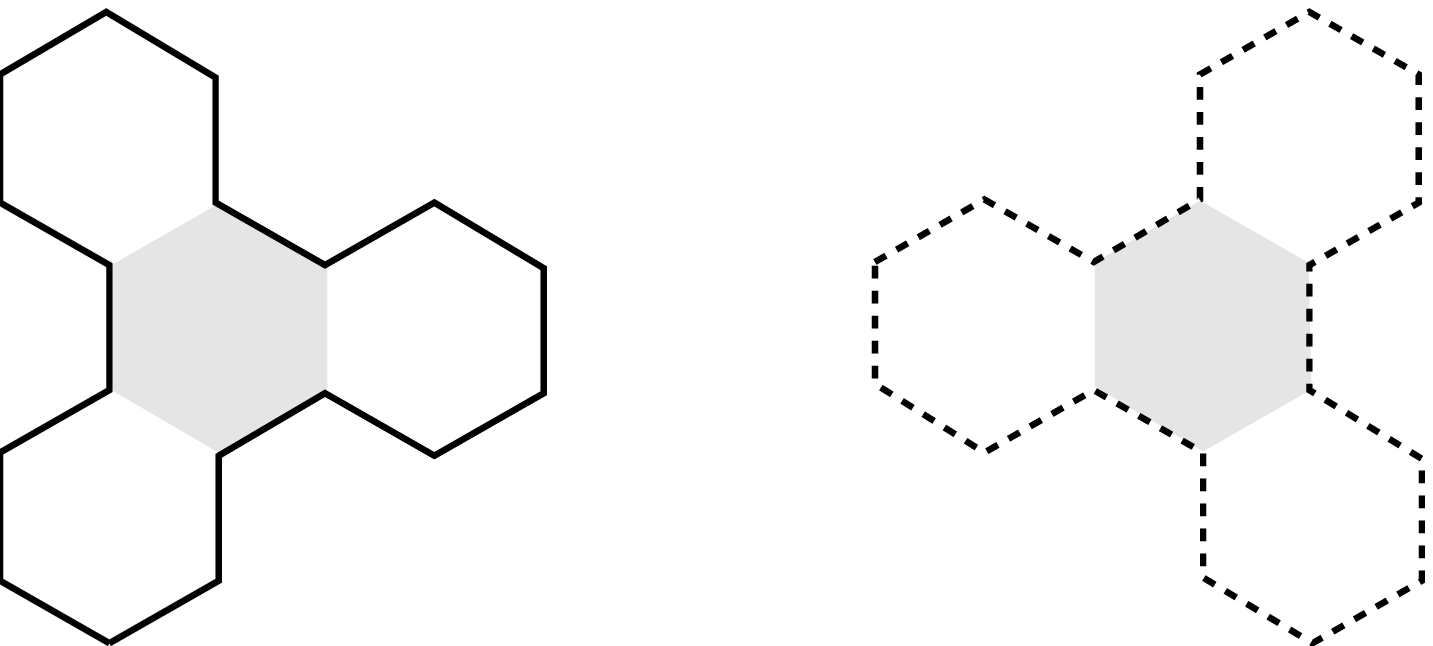}{5.cm}
\figlabel\excitation

Starting from the fundamental colouring state of Fig.\fundamental, 
we consider the following local excitation:
pick a particular $BR$ hexagon, and exchange its $B$ and $R$ links.  
The effect of a $BR$
excitation on the neighbouring $BW$ and $WR$ loops is illustrated in Fig.\excitation.
In this new colouring
configuration, $m_1$ is clearly unchanged, 
but the three $BW$ neighbouring hexagons have been glued into
just one loop (thick solid line on Fig.\excitation), hence $m_2$ is decreased by $2$, 
and the same applies to the three $WR$ neighbouring
hexagons (thick dashed line in Fig.\excitation), hence $m_3$ is also decreased by $2$. 
Therefore the total weight of the excited configuration is
\eqn\excited{ 2^{{N_\Delta \over 2} -4}. }
Now there may be any number of $BR$ local excitations in the system, but their 
interactions will result in a modification of their weight.  
The number of loops
can only be increased by such interactions however, as will become clear below. 
{}From now on, we factor out the term $\sqrt{2}^{N_{\Delta}}$ in the partition function.
The above leads to a lower bound
for $Z_{3d}$ by considering a non--interacting gas of $BR$ excitations with fugacity $1/2^4$, 
taking place among the $N_H/3$ hexagons of $BR$ type, 
with partition function
\eqn\lowgas{ Z^{(0)}_{BR}~=~ 2^{N_\Delta \over 2}  \times (1+{1 \over 2^4})^{N_H \over 3}~=~
\left({17 \over 2}\right)^{N_\Delta \over 6} \ .}
Hence the improved lower bound on $q_{3d}$
\eqn\improlo{ q_{3d}~\geq ~ \left({17 \over 2}\right)^{1 \over 6}~=~ 1.4285...}

\fig{Two neighbouring $BR$ excitations. 
The $BR$ excitations take place on the shaded hexagons. The first figure shows
how five former $BW$ hexagons have been glued into a single loop 
(thick solid line) leading to $m_2 \to m_2-4$. 
The second figure shows how five former $WR$ hexagons have been glued into
a single loop (thick dashed line) leading to $m_3 \to m_3 -4$.}{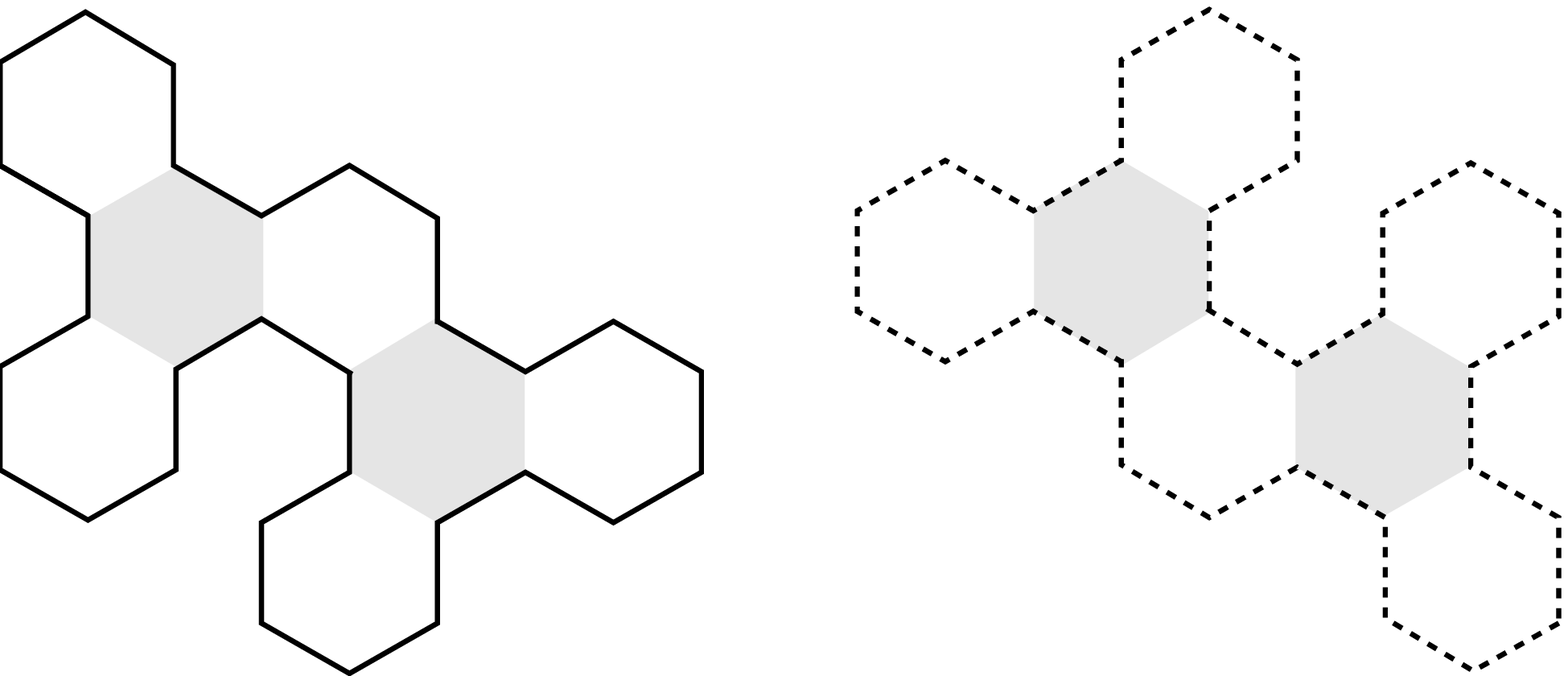}{8.truecm}
\figlabel\excitwo
The $BR$ excitations take place on the $BR$ hexagons of the fundamental state
of Fig.\fundamental, i.e. on the vertices of a triangular lattice of larger size, which we call
the excitation lattice from now on.
Excitations which are not neighbours on this lattice
do not interact, namely their total weight factorises into the product of their individual weights.
In fact, there is no two--body interaction between them: when two such excitations are neighbours
as illustrated in Fig.\excitwo,
i.e. only connected by a $W$ link, the total number of loops
is reduced by $8$ (five neighbouring $BW$ hexagons are glued into
one loop, hence $m_2 \to m_2-4$, and five neighbouring $WR$ hexagons are glued into one loop,
hence $m_3 \to m_3 -4$); hence a relative weight $1/2^8=1/2^4 \times 1/2^4$. 

\fig{Three neighbouring $BR$ excitations. The picture on the left shows how the $6$ 
former neighbouring $BW$ hexagons have been glued to form two loops 
(thick solid lines) leading to $m_2 \to m_2-4$. The picture on the right
shows how the $7$ neighbouring $WR$ hexagons have been glued to form a single 
loop (thick dashed line) leading to $m_3 \to m_3 -6$. The total number of loops
is therefore decreased by $10=12-2$, showing that the three--body interaction 
weight is $2^2$. Note the asymmetry between $BW$ and $WR$ loops.}{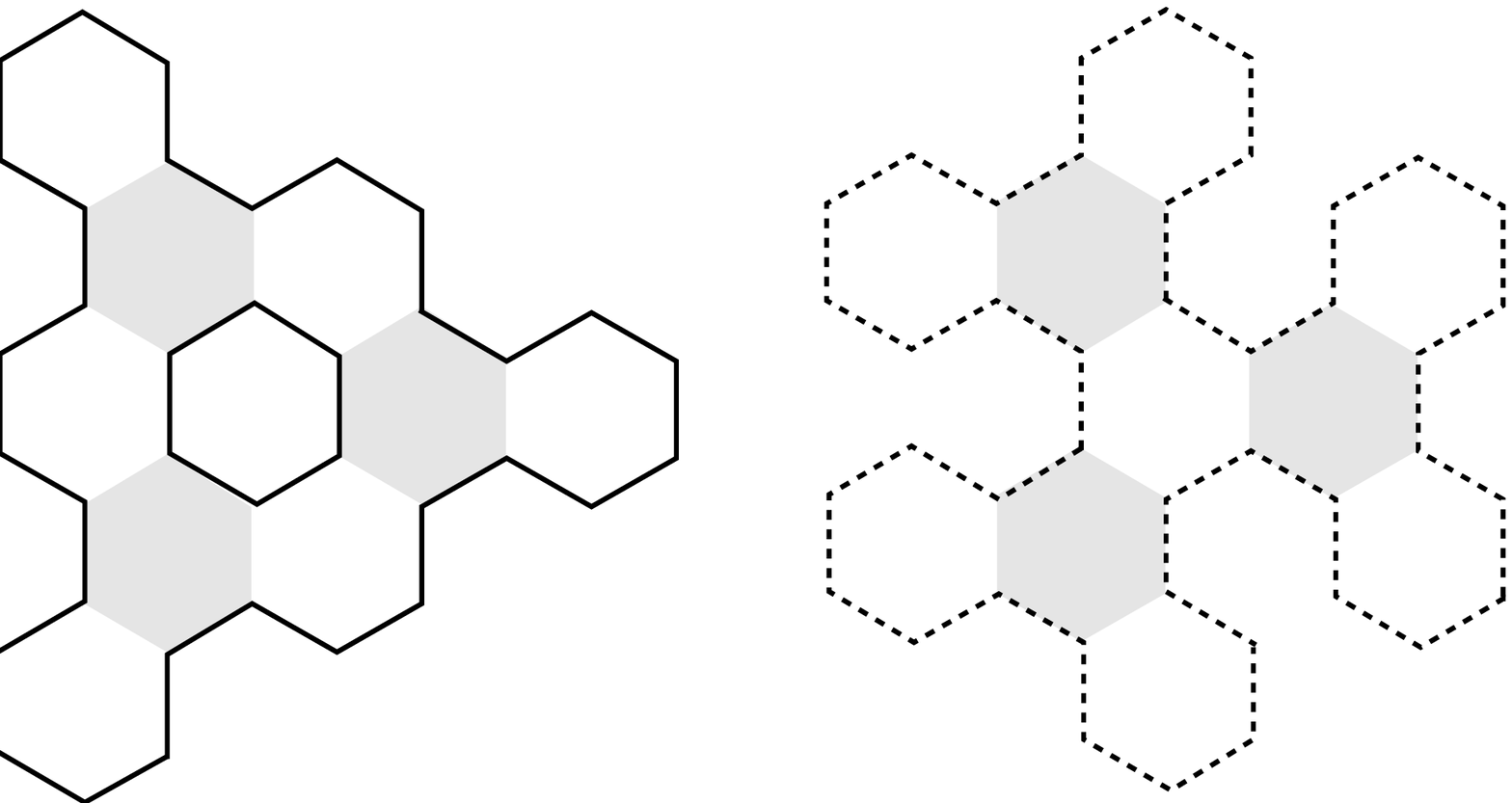}{8.truecm}
\figlabel\excithree
There is, however,
a non--trivial three--body interaction, when three $BR$ excitations take place on the three
$BR$ neighbours of a $BW$ or a $WR$ hexagon, as depicted in Fig.\excithree.  
Indeed, the three excitations re--create a central
loop with $B$ and $R$ exchanged, thus increasing the total number of loops. 
A more precise
counting (see Fig.\excithree)
shows in this case that the total number of loops is increased by $2$ (only $10$ loops have
disappeared, instead of $12=3\times 4$ for three independent excitations), resulting
in a three--body interaction weight $2^2$. The partition function $Z_{BR}$ incorporating
any $BR$ excitation can be evaluated order by order in the number of $BR$ excitations.
At this stage, the lower bound \improlo\ can therefore still be improved by performing a Mayer
expansion of the free energy per hexagon, in increasing order of
the number of $BR$ excitations. Up to order
$5$ in the number of $BR$ excitations, we find the partition function per 
hexagon
\eqn\betterq{ (q_{BR})^6~=~8
( 1 + {1 \over 2^4} + {3 \over 2^{11}}+{3 \over 2^{13}}+{69 \over 2^{19}}+\cdots) 
\geq  8\times 1.0644...}
hence the improved lower bound
\eqn\impropro{ q_{3d} ~\geq ~ q_{BR}~ \geq ~ 1.429...}

Let us finally show that the contribution 
$Z_{BR}$ to the partition function of the
most general combination of $BR$ excitations of the above groundstate is directly expressible
as the high temperature expansion of a particular $O(n)$ model on the hexagonal lattice.  
\fig{A typical cluster of $BR$ excitations. 
The $BR$ hexagons are represented by thin dotted lines: 
their centers sit at the vertices of a triangular lattice. 
The shaded hexagons are excited, namely their $B$ and $R$ links are exchanged. 
The two--component thick solid line is the boundary of the cluster, drawn on
the links of the original triangular lattice (thin dotted lines). The $3$ loops
in thick dashed lines result from the gluing of the $BW$ hexagons touching the
boundary of the cluster. This reduces number of $BW$ loops 
$m_2 \to m_2-10$. The $WR$ loops have not been represented for simplicity.
The reader will convince himself that their number is reduced by the cluster to
$m_3 \to m_3-12$. On the other hand, the two connected components
of the boundary have respective lengths $6$ and $20$, satisfying
$(6-2)+(20-2)=10+12$, as expected.}{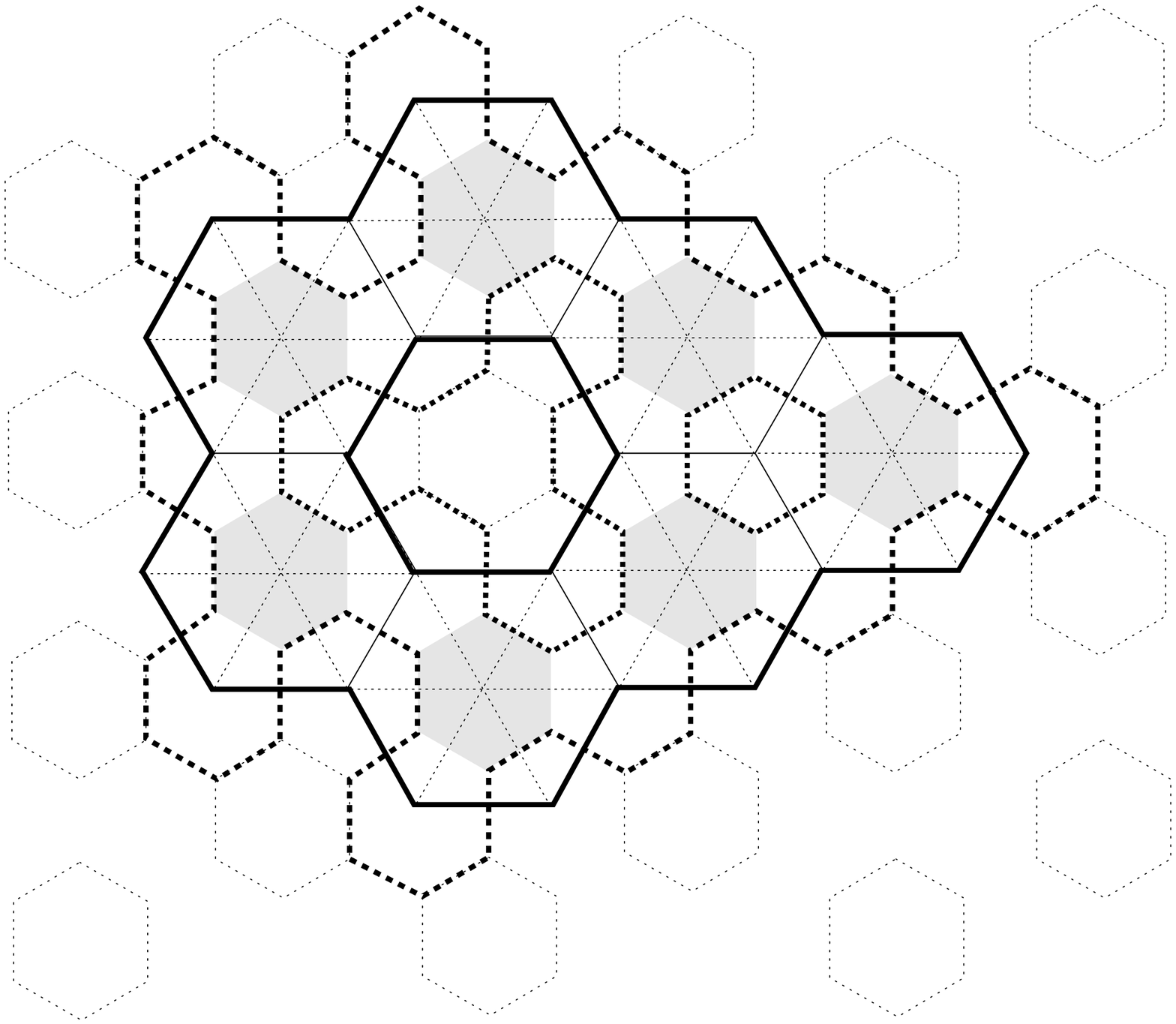}{8.truecm}
\figlabel\clus
\noindent Indeed, any combination of excitations can be decomposed into {\it clusters} which 
do not interact with each other.  
A typical cluster of $BR$ excitations is shown in Fig.\clus: the $BR$ hexagons are 
represented by thin dotted lines; the excited $BR$ hexagons 
(i.e. with $B$ and $R$ links
exchanged) forming the cluster are shaded. Note that the $W$ links stay in the 
initial state of Fig.\fundamental, as they are not affected by
$BR$ excitations. The clusters are indeed clusters of neighbouring vertices on the excitation lattice.
The dual white links form an 
hexagonal sublattice of the original triangular lattice, 
which is also the dual of the excitation lattice.
We define the {\it boundary} of a given 
cluster as its envelope of $W$ links on this lattice (the boundary does not include internal $W$ 
links which separate neighbouring $BR$ excitations in the cluster). 
The boundary of the cluster is represented in Fig.\clus\ by thick solid lines: note that the boundary 
links (which are $W$ links of the original triangular lattice) 
cut white links of former $BW$ or $WR$ hexagons, which are glued together by the $BR$ excitation.
A given cluster $C$ receives the relative weight
\eqn\cluw{ \prod_{{\rm boundary} \ {\rm connected} \atop
{\rm components}\ c} 2^{2-\ell_c} }
where the product extends over all the connected components $c$ of the boundary of the cluster
(there are two such components for the boundary of the cluster of Fig.\clus). The 
number $\ell_c$ denotes the total length of the connected component $c$, namely the total 
number of $W$ links
forming $c$ (the two corresponding lengths are $6$ and $20$ for the example of Fig.\clus). 

This result can be easily understood qualitatively. 
In the bulk inside a cluster, the hexagonal loops of $BW$ and $WR$ type are simply exchanged.
Therefore, the changes in the numbers of loops
can only take place at the boundary of the cluster, 
where some $BW$ (resp. $WR$) hexagons are glued together to form larger loops, 
thus decreasing the values of $m_2$ (resp. $m_3$).

Eq.\cluw\ can be proved by recursion, by studying the effect of adding
an extra $BR$ excitation to a given cluster $C$. Of course \cluw\ holds for a single isolated
$BR$ excitation, in which case $\ell_c=6$, and we recover the individual weight
$1/2^4$ \excited. 

Notice now that the weight \cluw\ only depends on the boundaries of the clusters, 
which form a set of
loops drawn on the white links of the groundstate of Fig.\fundamental.
This leads to a reexpression of
the partition function $Z_{BR}$ for all possible $BR$ excitations as that of a loop gas on this 
(original $W$ links) hexagonal lattice
\eqn\loopbr{ Z_{BR}~=~ 2^{N_\Delta \over 2} ~
\sum_{{\rm loops}\ {\rm on}\ {\rm the}\atop {\rm hexag.}\ {\rm latt.}}
~ \prod_{\rm loops} ~  2^{2-\ell}  }
where $\ell$ denotes the length of each loop.
The sum in eq.\loopbr\ is nothing but the high temperature expansion of the $O(n)$ model 
of \NIEN\ on the hexagonal lattice, with $n=4$ and $K=1/2$. Hence
\eqn\lopgas{ q_{BR}~=~ \sqrt{2}~\big(q_{O(n=4)}(K=1/2)\big)^{1 \over 6} }
where $q_{O(n)}(K)$ denotes the thermodynamic partition function per hexagon
of the $O(n)$ model.
Unfortunately the partition function for the $O(n=4)$ model is not known exactly. 
The model can be mapped \NIEN\ onto a $6$--vertex model on the Kagom\'e lattice,
whose vertices sit at the center of the links of the hexagonal lattice.  The latter
however is a $6$--vertex model {\it in a magnetic field}, for which no exact solution
is available.
Nevertheless, the
Mayer expansion above \betterq\ is expected to converge rapidly, and the value found for the 
lower bound should be accurate to the first three digits
(the order $5$ term in the expansion is only $\sim 10^{-4}$). 

Still $q_{BR}$ as given by \lopgas\ is only a lower bound on $q_{3d}$.

\listrefs

\end